\documentclass[12pt,a4paper,twoside,twocolumn,english,english]{article}
\usepackage{setspace}  
\usepackage{babel}     
\usepackage{calc}      
\usepackage[T1]{fontenc}       
\usepackage[utf8]{inputenc}  
\usepackage{ulem}
\usepackage{adjustbox}
\usepackage{graphicx}
\usepackage[font=footnotesize,labelfont=bf]{caption}
\usepackage[list=true, font=footnotesize, labelfont=bf, labelformat=parens, position=top]{subcaption}
\usepackage{here}

\usepackage{listings}
\usepackage{listingsutf8}
\usepackage{appendix}

\usepackage{amsmath,amssymb,amsfonts,amstext} 

\usepackage{lmodern} 
\usepackage[comma,numbers,sort&compress,super]{natbib}

\usepackage{booktabs}                      
\usepackage{longtable}                     
\usepackage{array}                         

\usepackage{textcomp,gensymb}

\newcommand{\beginsupplement}{%
        \setcounter{table}{0}
        \renewcommand{\thetable}{S\arabic{table}}%
        \setcounter{figure}{0}
        \renewcommand{\figurename}{Extended Data Figure}
     }

\usepackage{units}   

\usepackage[pdfstartview=FitH,      
            breaklinks=true,        
            bookmarksopen=true,     
            bookmarksnumbered=true  
            ]{hyperref}
\usepackage{xcolor}
\hypersetup{
    colorlinks,
    linkcolor={black},
    citecolor={black},
    urlcolor={black}
}

\usepackage{epsf}
\usepackage{authblk} 

\title{Interstitial segregation has the potential to mitigate liquid metal embrittlement in iron}
\author[1]{Ahmadian, A.}
\author[2]{Scheiber, D.}
\author[1]{Zhou, X.}
\author[1,3]{Gault, B.}
\author[2,4,*]{Romaner, L.}
\author[5]{Darvishi Kamachali, R.}
\author[2]{Ecker, W.}
\author[1,*]{Dehm, G.}
\author[1,*]{Liebscher, C. H.}
\affil[1]{Max-Planck-Institut fuer Eisenforschung GmbH, 40237 Düsseldorf, Germany}
\affil[2]{Materials Center Leoben GmbH, 8700 Leoben, Austria}
\affil[3]{Department of Materials, Royal School of Mines, Imperial College London, London, UK}
\affil[4]{Montanuniversität Leoben, Leoben, Austria}
\affil[5]{Federal Institute for Materials Research and Testing (BAM), Unter den Eichen 87, 12205 Berlin, Germany}
\affil[*]{\textit {Corresponding authors: dehm@mpie.de, lorenz.romaner@unileoben.ac.at, c.liebscher@mpie.de}}
\date{}                     
\begin{document}

\maketitle

\begin{otherlanguage}{english}
\begin{abstract}
The embrittlement of metallic alloys by liquid metals leads to catastrophic material failure and severely impacts their structural integrity. The weakening of grain boundaries by the ingress of liquid metal and preceding segregation in the solid are thought to promote early fracture. However, the potential of balancing between the segregation of cohesion-enhancing interstitial solutes and embrittling elements inducing grain boundary decohesion is not understood. Here, we unveil the mechanisms of how boron segregation mitigates the detrimental effects of the prime embrittler, zinc, in a $\Sigma 5\,[0\,0\,1]$ tilt grain boundary in $\alpha-$Fe ($4~at.\%$ Al). Zinc forms nanoscale segregation patterns inducing structurally and compositionally complex grain boundary states. Ab-initio simulations reveal that boron hinders zinc segregation and compensates for the zinc induced loss in grain boundary cohesion. Our work sheds new light on how interstitial solutes intimately modify grain boundaries, thereby opening pathways to use them as dopants for preventing disastrous material failure.  
\end{abstract}

\section*{Introduction}
Grain boundary (GB) properties such as cohesive strength or mobility can be significantly altered through segregation of alloying elements or impurities \cite{lejcekAnalysisSegregationinducedChanges2014}. One of the most detrimental elements impacting the properties of GBs in iron (Fe) and steel is zinc (Zn), which promotes brittle intergranular fracture of the material through liquid metal induced embrittlement (LMIE) \cite{ hillertChemicallyInducedGrain1978, bealLiquidZincEmbrittlement2012, choMicrostructureLiquidMetal2014, ashiriSupercriticalAreaCritical2015, razmpooshLiquidMetalEmbrittlement2018, razmpooshRoleRandomCoincidence2020b, ikeda2022early,scheiberInfluenceAlloyingZn2020a}.


Several models were proposed to explain the underlying mechanisms leading to LMIE \cite{fernandesMechanismsLiquidMetal1997}. The Stoloff–Johnson–Westwood–Kamdar (SJWK) model considers that the liquid metal adsorbed at the crack tip promotes crack propagation under an applied stress by weakening the strength of the interatomic bonds \cite{stoloffCrackPropagationLiquid1963, westwoodConcerningLiquidMetal1963}. In contrast, stress-assisted GB diffusion preceding the crack initiation is assumed to be the root cause for the increase in grain boundary brittleness in the Krishtal-Gordon-An model\cite{gordonMechanismsCrackInitiation1982}. DiGiovanni et.~al \cite{digiovanniLiquidMetalEmbrittlement2021} investigated the diffusion of Zn in Fe by means of electron-probe microanalysis (EPMA) before and after mechanical loading. Their observations confirmed that stress-assisted diffusion in the vicinity of the crack tip is one of the dominant mechanisms. There are various explanations on the origin of the segregation induced GB embrittlement process. Atomic size differences in Cu-Bi, also an LMIE system, were shown to cause local strain fields, which facilitate a reduction in GB cohesion \cite{schweinfestBismuthEmbrittlementCopper2004}. However, the size differences between Fe and Zn are not large enough to play a major role. Instead electronic effects such as bonding type \cite{senelLiquidMetalEmbrittlement2014} and mobility \cite{yuasaBondMobilityMechanism2010, wuFirstPrinciplesDetermination1994, Wu1994} can be made responsible for GB embrittlement. Peng et~al. \cite{pengEffectZincdopingTensile2020} performed first-principles tensile tests on a Zn segregated $\Sigma 5\,[0\,0\,1]$ GB in  ferritic Fe. They found a weakening of the GB because of covalent bonding between Zn and Fe, which ultimately reduces the charge density between the Fe atoms and hence their bond strength.

In order to prevent LMIE, it was proposed to introduce cohesion enhancing alloying elements to GBs. Scheiber et~al. \cite{scheiberInfluenceAlloyingZn2020a} concluded from ab-initio and thermodynamic modelling that the embrittling effect of Zn at different symmetrical bcc-Fe tilt GBs can be reduced by Al and Si, which is related to site-competition as well as repulsive interactions between the solutes and Zn. Although a large body of experimental and theoretical studies on LMIE exist \cite{gengInfluenceAlloyingAdditions2001, razmpoosh2021pathway, bhattacharyaInfluenceStartingMicrostructure2021}, a fundamental understanding on the relationships between the GB structure, Zn segregation and its impact on embrittlement is lacking. Especially, the role of interstitial solutes on the embrittlement of GBs by Zn has not been considered, while for example C and B are known to act as GB cohesion enhancers in Fe-based alloys \cite{ahmadianAluminumDepletionInduced2021, wuEffectsCarbonFeGrainBoundary1996, fraczkiewiczInfluenceBoronMechanical2000, wangFirstPrinciplesStudyCarbon2016, miyazawaAtomicBondbreakingBehaviour2017,kholtobina2021effect}. Miyazawa et~al. \cite{ miyazawaAtomicBondbreakingBehaviour2017} have shown by first-principles based calculations that segregation of C would increase the GB cohesion in Fe. In contrast to C, B can occupy substitutional as well as interstitial positions in $\alpha$-Fe \cite{forsNatureBoronSolution2008} and would enhance the cohesion strength of the GB when taking up interstitial positions \cite{wuFirstPrinciplesDetermination1994}. Therefore, it is of great interest to understand how cohesion enhancing solutes such as B and C impact the segregation behavior of Zn and how their interplay affects the detrimental effect of Zn on GB cohesion on an atomistic level. 

Here, we study the local atomic structure and composition of a $\Sigma 5$\,(3\,1\,0)[0\,0\,1] tilt GB in a bicrystal of body centered cubic (bcc) Fe at different levels of Zn segregation. By atomically resolved microscopic probing techniques, we find that Zn forms nanometer sized segregation lines in a region preceding the former solid-liquid interface. However, both B and C are homogeneously distributed across the GB plane. The Zn-segregation pattern formation is explained by density-based phase-field simulations revealing a miscibility gap of the GB, which promotes a GB phase decomposition. Furthermore, our ab-initio calculations show that the presence of B and C significantly reduces the segregation energy of Zn due to strong repulsive interactions, which also leads to a reduction of the maximum Zn concentration at the GB. Interestingly, B is capable to compensate for the reduction in the work of separation through Zn even for high Zn GB concentrations. Our work suggests that interstital solutes such as B and C have the potential to mitigate the embrittling tendency of Zn. These insights provide pathways to prevent liquid metal embrittlement in Fe by controlled addition of cohesion enhancing elements. 

\section*{Results}
\noindent{\textbf{Global grain boundary structure}}\\

\vspace{-1em}
We use a bicrystal containing a $\Sigma 5$\,(3\,1\,0)[0\,0\,1] tilt GB with trace amounts of B and C to study Zn segregation from the liquid state into the GB and the interaction of Zn with the interstitial solutes. We formed a diffusion couple of Zn and the bicrystal, and subsequently annealed it at $800^\circ$C to liquefy the Zn reservoir and enable diffusion of Zn into the solid Fe bicrystal as schematically shown in Fig.~\ref{pub2_SEM}a. The detailed experimental procedure is described in the Methods and Extended Data sections. The backscattered electron (BSE) secondary electron microscopy (SEM) image in Fig.~\ref{pub2_SEM}b shows a two-phase region at the interface between the former liquid Zn reservoir and the Fe bicrystal consisting of $\alpha-$Fe (bcc) and $\Gamma-$Zn (see Extended Data). Here, we limit our observations to the GB below this two-phase region. We specifically focus on two different GB areas indicated as "Region $1$", which is Zn-rich with an average concentration of Zn determined by energy-dispersive X-ray spectroscopy (EDS) in the SEM of $\sim$5~at.$\%$, and "Region $2$" with an average Zn content of $\sim$0.5~at.$\%$ (see Fig.~\ref{pub2_SEM}b
). Region $1$ is $\sim$5~$\mu m$ below the two-phase region and high angle annular dark-field (HAADF) imaging along the $[0\,0\,1]$ tilt axis in the scanning transmission electron microscope (STEM) reveals that the GB inclination is locally deviating from the exact (310) habit plane (see Fig.~\ref{pub2_SEM}c). Corresponding STEM-EDS measurements shown in Fig.~\ref{pub2_SEM}d demonstrate strong segregation of Zn to the GB and suggest that it is inhomogeneously distributed along the GB. The GB in Region $2$, which is $\sim$30$\,\mu m$ away from the two-phase region, is curved and exhibits a high density of local kinks as shown in Fig.~\ref{pub2_SEM}e. Zinc is only slightly enriched at the GB as shown in the elemental map in Fig.~\ref{pub2_SEM}f.
\begin{figure*}[hbtp]
  \centering
\includegraphics[width=0.98\linewidth]{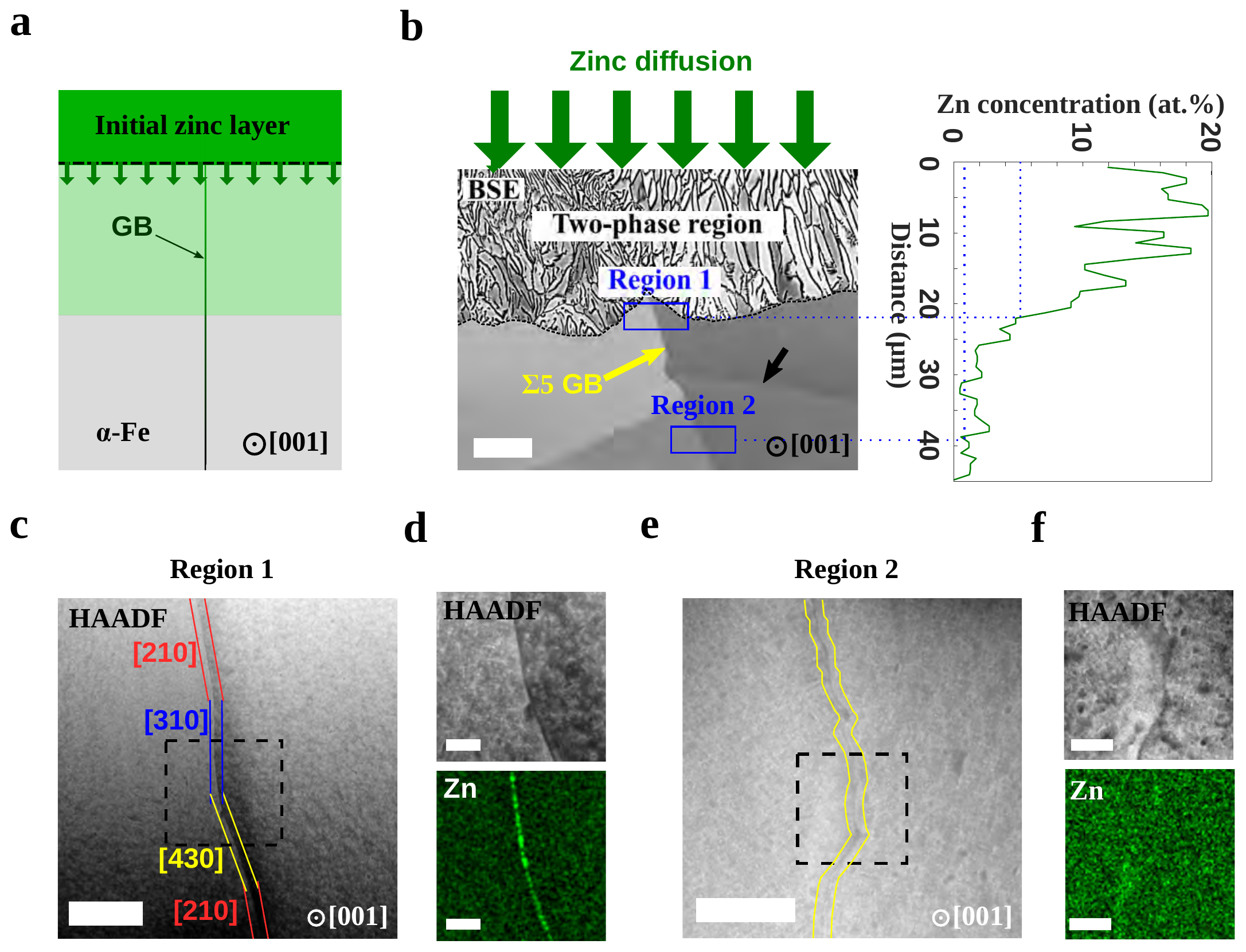}
\caption{\textbf{Overview BSE images of the $\Sigma 5\,[0\,0\,1]$ GB. a-b} The diffusion of Zn results in the formation of a two-phase region containing $\alpha$-Fe and $\Gamma-$ Fe-Zn phases. STEM specimen were extracted from two regions (indicated by blue rectangles). \textbf{b} Region 1 marks an area of the GB in the vicinity of the two-phase region, while Region 2 is $\sim 30\,\mu m$ away from the two-phase region. Furthermore \textbf{b} shows a low-angle GB (indicated by a black arrow) intersecting with the $\Sigma$5 GB and locally bending it. \textbf{c-d} Region $1$: \textbf{c} HAADF-STEM image of the $\Sigma 5$ GB viewed along the $[0\,0\,1]$ tilt axis. The GB is inclined and the boundary plane can be divided into nearly symmetric $(2\,1\,0)$, $(4\,3\,0)$ and $(3\,1\,0)$ segments. \textbf{d} STEM-EDS mapping of the region, which is indicated in \textbf{c} by a black dashed rectangle. The GB is enriched by Zn with a weak decrease from top to bottom. \textbf{e-f} Region $2$: \textbf{e} HAADF-STEM image shows strong inclinations of the boundary. \textbf{f} Corresponding STEM-EDS map  of the black dashed rectangle region shows no clear enrichment of Zn. The scale bar in \textbf{b} is $10\,\mu m$, in \textbf{c} it is $100\,nm$, in \textbf{d}, $20\,nm$, in \textbf{e} $50\,nm$ and in \textbf{f} $20\,nm$}.
\label{pub2_SEM}
\end{figure*}

\vspace{1em}
\noindent{\textbf{Near atomic scale zinc segregation}}\\

\vspace{-1em}
We performed near atomic resolution STEM-EDS elemental mapping in the Zn-rich Region $1$ of the bicrystal. Figure~\ref{pub2_HRSTEM-EDS} shows atomic resolution STEM images and the associated Zn distribution along the GB at acceleration voltages of $300\,kV$ and $120\,kV$, respectively. The HAADF-STEM images in Fig.~\ref{pub2_HRSTEM-EDS}a and b reveal that the GB structure appears disrupted by a high density of GB defects. Measurements at $120\,kV$ shown in Fig.~\ref{pub2_HRSTEM-EDS}b clearly demonstrate clustering of Zn at the GB into Zn-rich ($\sim$35~$at.\%$) and Zn-lean ($\sim$10~$at.\%$) regions having a width of $\sim$1~nm and a regular spacing of $\sim$3-4~nm. Although Zn has a slightly higher atomic number than Fe ($\Delta$Z=4), the corresponding HAADF-STEM images do not show any indication of regions with higher intensity. More information on the STEM-EDS measurements will be discussed later. 

To obtain further insights into the 3D arrangement of the Zn segregation and its correlation to the distribution of the impurity elements B and C, we performed atom probe tomography (APT) experiments from the same regions as our STEM investigations.
\begin{figure*}[hbtp]
  \centering
\includegraphics[width=0.98\linewidth]{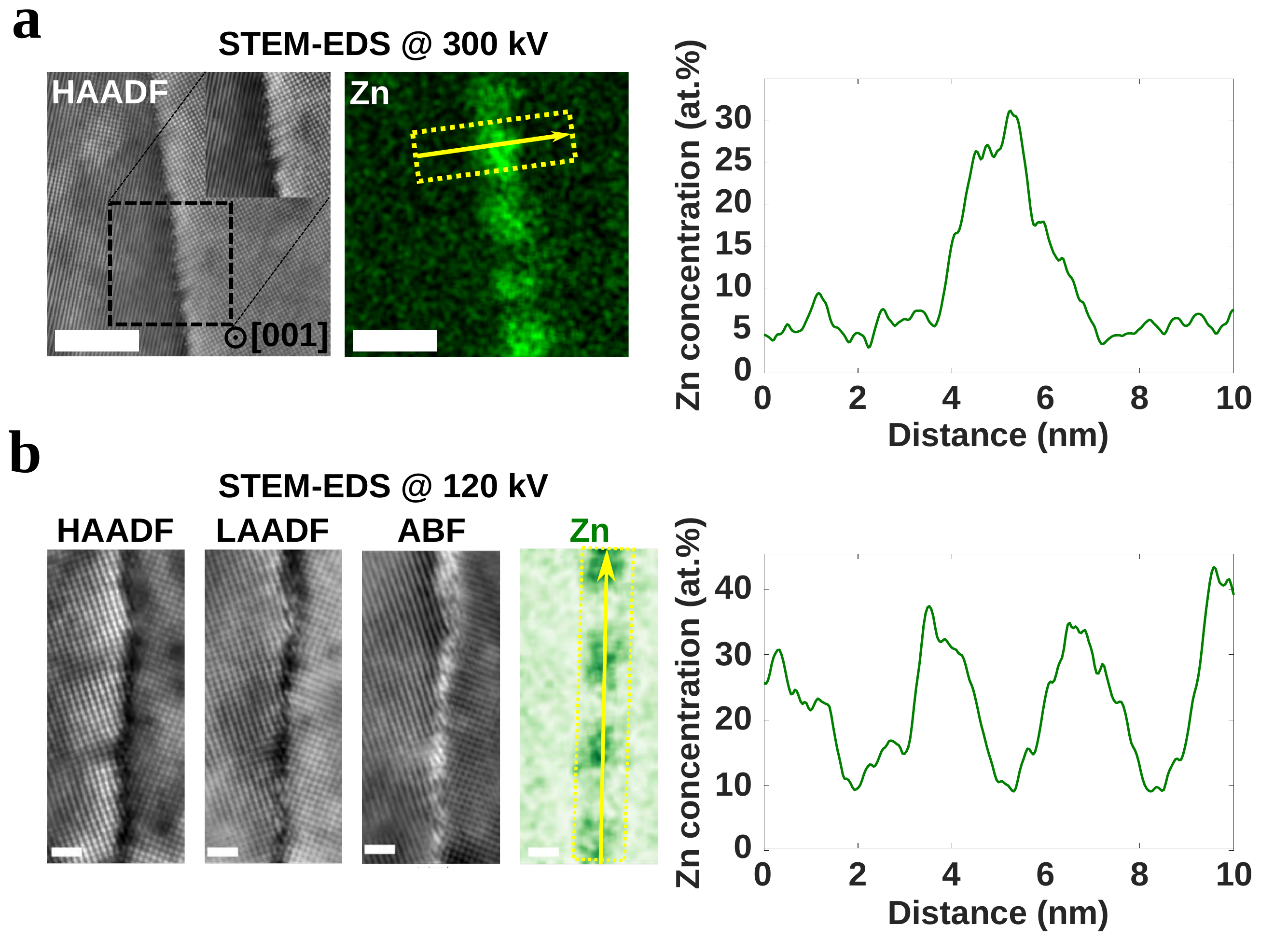}
\caption{\textbf{High resolution STEM-EDS at different acceleration voltages. a} The HAADF-STEM and corresponding Zn elemental map show formation of Zn clusters with concentrations of $\sim 30\,at.\%$.  To improve the STEM-EDS resolution, the acceleration voltage was reduced to $120\,kV$ as shown in \textbf{b}. The incident electrons are scattered to lower angles due to the presence of  the Zn clusters resulting in a dark contrast in the HAADF- as well as LAADF-STEM image but a bright contrast in the ABF-STEM image. The Zn concentration profile along the GB shows a periodicity of $3-4\,nm$ with concentrations varying from $10\,at.\%$ to $40\,at.\%$. The scale bar in \textbf{a} is $5\,nm$ and in \textbf{b} $1\,nm$.}
\label{pub2_HRSTEM-EDS}
\end{figure*}

\vspace{1em}
\noindent{\textbf{Grain boundary composition}}\\

\vspace{-1em}
Figure~\ref{pub2_APT}a presents the reconstructed $3$D-volume of the needle-shaped specimen viewed along the $[0\,0\,1]$ tilt axis extracted from Region $1$ (Zn-rich), where the GB is positioned at a 45$^{\circ}$ angle $\sim$100~nm away from the apex. Besides the segregation of Zn, B and C are also observed to segregate to the GB. The 3D atom map of B shows a rather unusual behavior in the form of a tail that extends into the lower grain for nearly 50~nm, although it has been recognized that anomalies in the field evaporation of solutes can lead to asymmetric solute distribution at GBs \cite{felferNewApproachDetermination2012}. The concentration profile of B extracted across the GB in Fig.~\ref{pub2_APT}b verifies this observation, which will be discussed later. The bulk Zn concentration of $\sim$5.5~$at.\%$ from APT agrees with STEM-EDS measurements (Fig.~\ref{pub2_HRSTEM-EDS}). The averaged GB concentration of Zn reaches a value of $\sim$10~$at.\%$. As we have seen before, the Al concentration decreases to $\sim$2.7$at.\%$ at the GB due to the presence of B and C \cite{ahmadianAluminumDepletionInduced2021}. Both concentration profiles of B and C show a broadened and asymmetric shape with a tail towards the left side corresponding to the lower grain in the reconstruction (Fig.~\ref{pub2_APT}a). The peak concentration of B and C in the initially investigated GB before Zn segregation was $1.8\,at.\%$ and $2.4\,at.\%$ \cite{ahmadianAluminumDepletionInduced2021}, which is reduced by a factor of $\sim$4 and $\sim$10, respectively. A close inspection of the peak locations of the concentration profiles reveals an offset of the B peak of $\sim$1~nm with respect to that of the Zn concentration peak. A similar behavior was obtained in a different APT specimen as shown in the Extended Data Figure~\ref{suppFig3}. 
\begin{figure*}[htbp]
\includegraphics[width=0.98\linewidth]{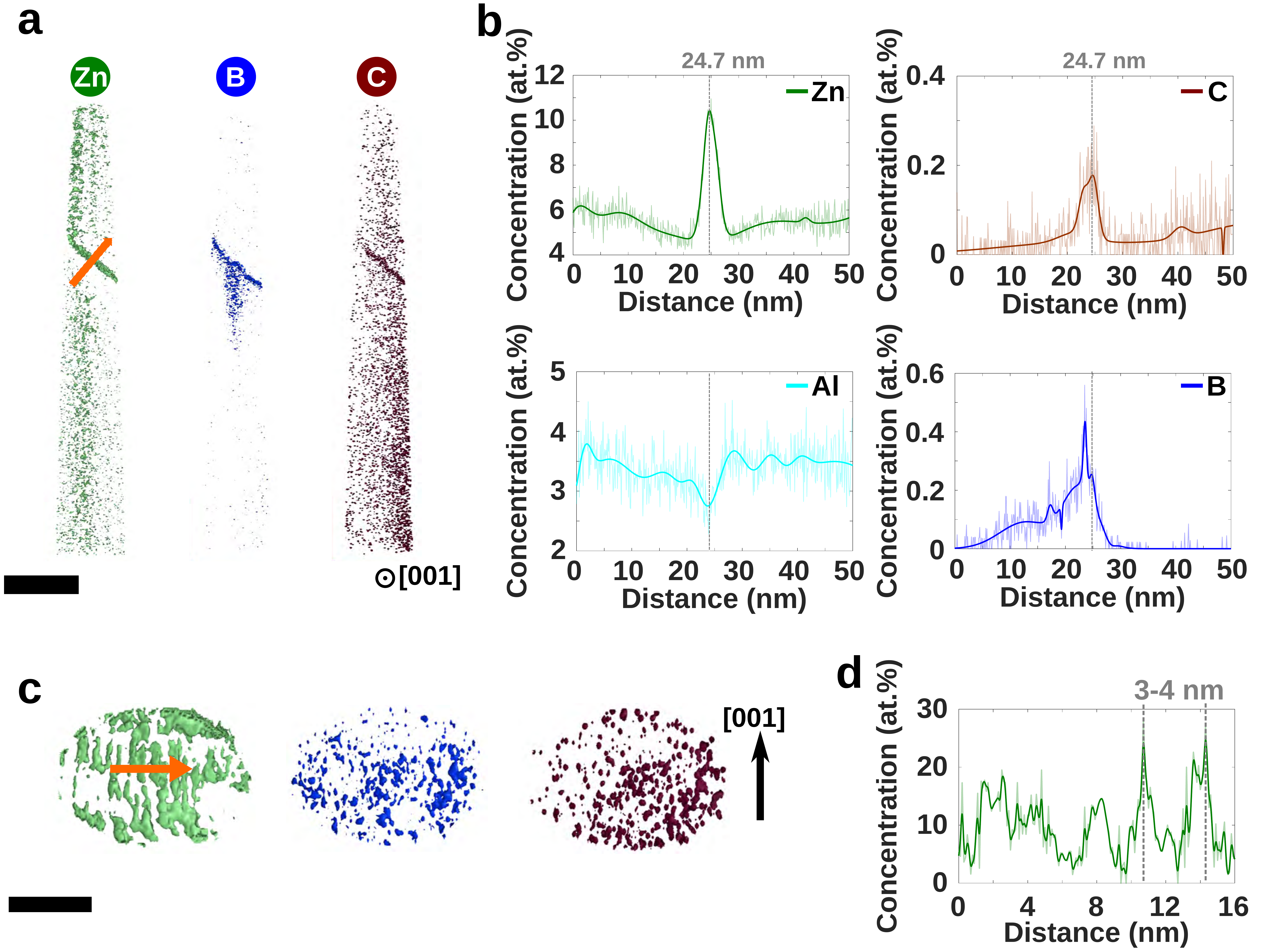}
\caption{\textbf{Segregation of Zn, B, C, and depletion of Al at the $\Sigma 5\, [0\,0\,1]$ GB of Region $1$}. \textbf{a} $3$D APT reconstruction showing the distribution of Zn, B and C viewed along the $[0\,0\,1]$ tilt axis. Using an isoconcentration value of $10\,at.\%$ for Zn, $0.5\,at.\%$ for B and $0.3\,at.\%$ for C highlights the segregation of these elements to the GB. \textbf{b} The composition profile extracted from a cylindrical region with diameter of $30\,nm$ is extracted across the GB (shown in \textbf{a} as orange arrow). A clear increase of Zn, B and C is shown at  the GB, while the concentration of Al decreases. The maximum concentration peak of Zn and C is at $24.7\,nm$, while B concentration reaches the maximum already at $23.5\,nm$ same as the minimum of Al. The Zn concentration first decreases slightly at the vicinity of the boundary until it increases steeply. The C and B concentration do not show such decrease at the vicinity of the GB. The B concentration shows a continuous increase over a large range of several tens of nm before reaching the maximum value and then it decreases abruptly to nearly $0\,at.\%$. \textbf{c} Rotation of the APT reconstruction in \textbf{a} such that the distribution of Zn, B and C onto the GB plane is shown, i.e. the viewing direction is along the GB plane. While B and C show no formation of patches, Zn forms columnar patches elongated along the tilt axis. \textbf{d} A concentration profile extracted along the orange arrow in \textbf{c} shows the concentration and distances of Zn columns. The scale bar in \textbf{a} is $50\,\text{nm}$ and $20\,\text{nm}$ in \textbf{c}.}
\label{pub2_APT}
\end{figure*}

Observations of the in-plane distribution of Zn within the GB obtained by APT reveals that Zn is arranged in the form of segregation lines which are aligned along the [001] tilt axis (see Figs.~\ref{pub2_APT}c and d). However, B and C show a homogeneous distribution within the GB plane as shown in Fig.~\ref{pub2_APT}c. A similar modulation in Zn concentration was observed by STEM-EDS shown in Fig.~\ref{pub2_HRSTEM-EDS}. 

APT investigations were also performed in Region $2$ (see Extended Figure~\ref{pub2_APT2}), revealing a lower average bulk Zn content of $\sim$0.5~$at.\%$. Interestingly, no clear indication for Zn segregation to the GB containing B and C  with peak concentration values of $\sim$1.8~$at.\%$ and $\sim$0.8~$at.\%$, respectively (shown in Extended Figure~\ref{pub2_APT2}b), is found. In comparison to the Zn-rich Region $1$ of Fig.~\ref{pub2_APT}b, the peak B and C concentration at the GB in the Zn-lean Region $2$ is nearly a factor of 4 higher, similar to values obtained in the  as-grown bicrystal \cite{ahmadianAluminumDepletionInduced2021}. Furthermore, the concentration profiles of B and C adopt a symmetric shape, in contrast to that observed in Region $1$ (Fig.~\ref{pub2_APT}b).

\vspace{1em}
\noindent{\textbf{Atomic grain boundary structure}}\\

\vspace{-1em}
To explore the underlying atomic GB structure in both Region $1$ (Zn-rich) and $2$ (Zn-lean), we used atomic resolution HAADF-STEM imaging as shown in Fig.~\ref{pub2_HRSTEM}. The HAADF-STEM overview image of the GB in Region $1$ shown in the Extended Figure~\ref{suppFig3}~a reveals a slight curvature of the GB, which leads to the formation of different nanoscale GB facets with varying GB planes close to $(3\,1\,0)$. The atomic resolution images of Fig.~\ref{pub2_HRSTEM}~a and b show that the GB structure is composed of kite-type structural units, which are disrupted by GB defects. Locally, the GB habit plane is varying between $(3\,1\,0)$- and $(2\,1\,0)$-type as well as asymmetric GB segments. 

The GB in Region $2$ (Zn-lean) is composed of nanofacets with different GB plane inclinations (see Fig.~\ref{pub2_SEM}~e and Supplementary Figure~\ref{suppFig3}~b). Figure~\ref{suppFig3}~b shows a magnified HAADF-STEM image of the GB in Region $2$, where the GB alternates between symmetric $(2\,1\,0)$ and asymmetric segments. The atomic GB structure in Fig.~\ref{pub2_HRSTEM}~c shows a near $(3\,1\,0)$ GB with kite-type structural units, which are interrupted by GB steps or disconnections (highlighted by dashed cyan circles) to compensate for deviations in GB inclination. In other areas in Region $2$ the GB is dissociating into a nanofaceted structure (see Fig.~\ref{pub2_HRSTEM}~d).

\begin{figure*}[htbp]
\includegraphics[width=0.98\linewidth]{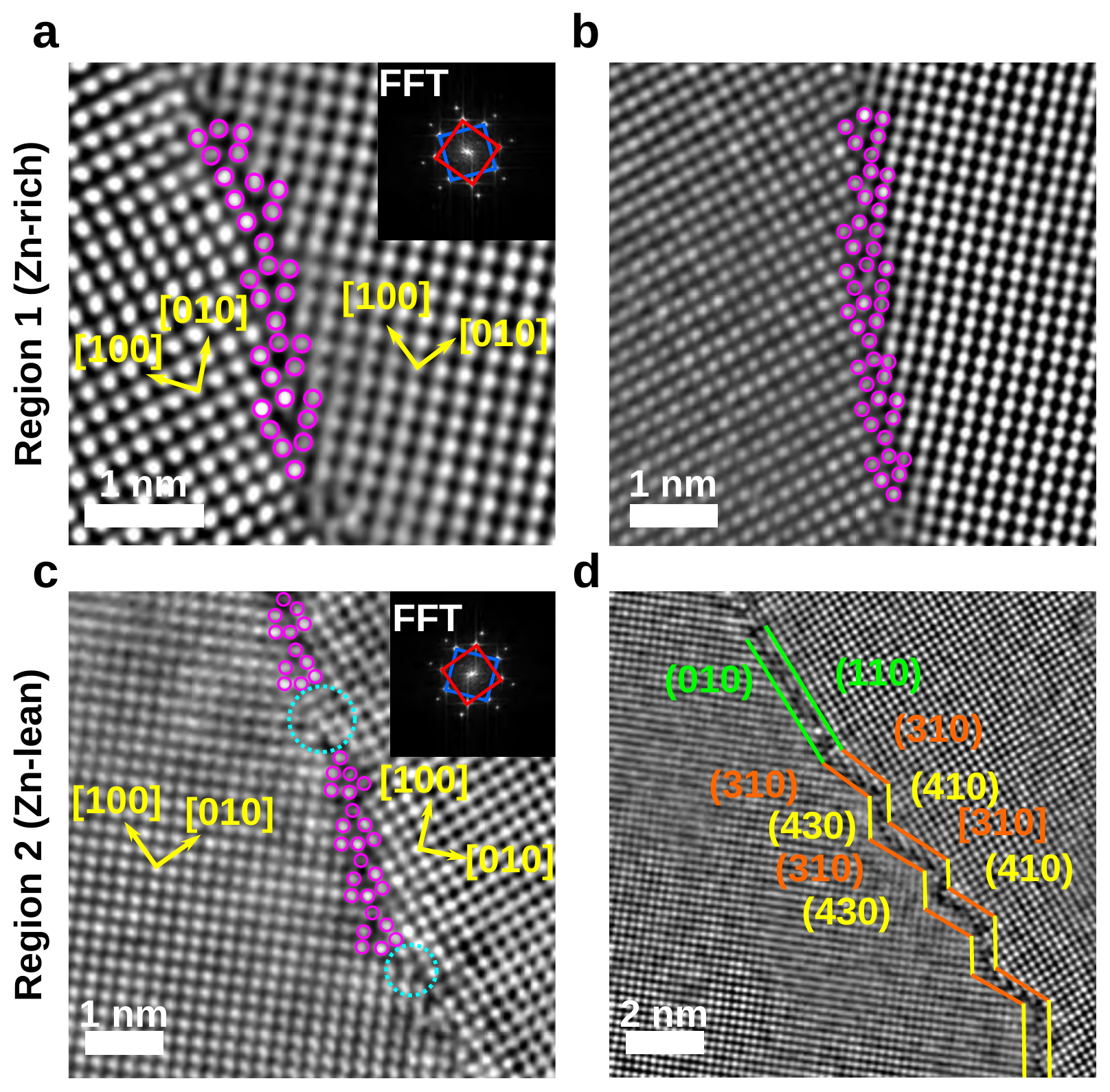}
\caption{\textbf{Atomic structure of the $\Sigma 5\,[0\,0\,1]$ GB in the Fe-4at.$\%$Al bi-crystal.} \textbf{a, b} High-resolution HAADF-STEM images from Region 1 (Zn-rich) shows the formation of kite-type structural units (purple color), where extra atoms are introduced to shift the SU parallel and perpendicular to the boundary plane normal. From the FFT in \textbf{a} the misorientation angle is determined as $40^\circ$. \textbf{c} High-resolution HAADF-STEM images from Region 2 (Zn-lean) of the area marked by the upper dashed square in Extended Figure~\ref{suppFig3}~b. The GB consists of perfect kite-type SU interrupted by GB defects, which result in the inclination of the boundary. \textbf{d} High-resolution HAADF-STEM image of the region marked by the lower dashed square in \ref{suppFig3}~b. The GB is faceting into symmetric $(3\,1\,0)/(3\,1\,0)$ and asymmetric  $(4\,3\,0)/(4\,1\,0)$. The scale bar in \textbf{a}, \textbf{b} and \textbf{c} marks $1\,\text{nm}$ and in \textbf{d} $2\,nm$.}
\label{pub2_HRSTEM}
\end{figure*}

\vspace{1em}
\noindent{\textbf{Grain boundary decomposition}}\\

\vspace{-1em}

The phase decomposition behavior of the $\Sigma 5\,(3\,1\,0)[0\,0\,1]$ GB was further explored by density-based phase-field (DPF) modeling and simulation. The computational details are given in the Methods section. In a first step, we modelled the GB phase diagram in the Fe-Zn system, which is shown in Fig.~\ref{PF_SIM}~a. Following this thermodynamic assessment, the GB exhibits a miscibility gap, which separates Zn-lean and Zn-rich phase regions, comparable to the Fe-Zn bulk phase diagram (see Extended Data Figure~\ref{DPF_bulk}). The origin of this phase separation in the $\alpha$-Fe-Zn system is related to magnetic ordering \cite{su2001thermodynamic}. The transition curve (blue) in Fig.~\ref{PF_SIM}~a represents the critical temperature dependent bulk concentration at which the $\Sigma 5\,(3\,1\,0)$ GB is expected to decompose into Zn-rich and Zn-lean domains. This also means that crossing the blue curve from left to right, e.g., via a change in temperature, a low-to-high Zn segregation transition is expected within the GB.

Interestingly, for a bulk Zn concentration of 5 at.\% as observed in Region 1 in the experiment (see Fig.~\ref{pub2_SEM}~b and~\ref{pub2_HRSTEM-EDS}~a) the DPF modeling predicts a GB phase decomposition at $\sim$403$^\circ$C (highlighted by the horizontal line in Fig.~\ref{PF_SIM}~a) and the GB is expected to decompose into a two-phase structure with Zn concentrations of $\sim17\,at.\%$ (Zn-lean) and $\sim52\,at.\%$ (Zn-rich), respectively. With decreasing temperature, the Zn-rich domains further enrich in Zn up to $\sim64\,at.\%$ at 275$^\circ$C and the concentration of the Zn-lean regions reduces slightly to $\sim 12\,at.\%$ (shown by the dashed horizontal line). The time dependent GB phase evolution was investigated by 3D DPF simulations at 300$^\circ$C and a bulk Zn concentration of 5~at.\% as shown in Fig.~\ref{PF_SIM}~b and c. Figure~\ref{PF_SIM}~b shows the evolution of the Zn concentration in an edge on view at the GB for four different time steps. The in-plane evolution of the Zn-rich and -lean domains are shown in Extended Data Figure~\ref{suppFig_DPF2}. Initially, Zn is homogeneously distributed along the GB with a concentration of $\sim 9\,at.\%$, as also seen in the concentration profiles extracted along the GB shown in Figure \ref{PF_SIM}~b (green curve). With progressing time, the GB phase separates into Zn-rich ($\sim 60\,at.\%$) and Zn-lean ($\sim 17\,at.\%$) domains, which are separated by $\sim5\,nm$ consistent with the experimental observations. For a bulk Zn concentration of 0.5 at.\%, no GB decomposition is observed according to the DPF simulations.

The predicted composition of the Zn-rich GB regions is higher than determined experimentally, since the GB phase decomposition is kinetically limited at lower temperatures in the experiment. This is evident from the APT measurements shown in Fig. \ref{pub2_APT}~b, where depletion of Zn adjacent to the GB is indicating incomplete segregation due to limited diffusivity during the cooling process. Furthermore, the presence of B and C atoms at the GB may impact the segregation behavior of Zn to the GB and ultimately affect the Zn GB concentration.

\begin{figure*}[htbp]
\includegraphics[width=0.98\linewidth]{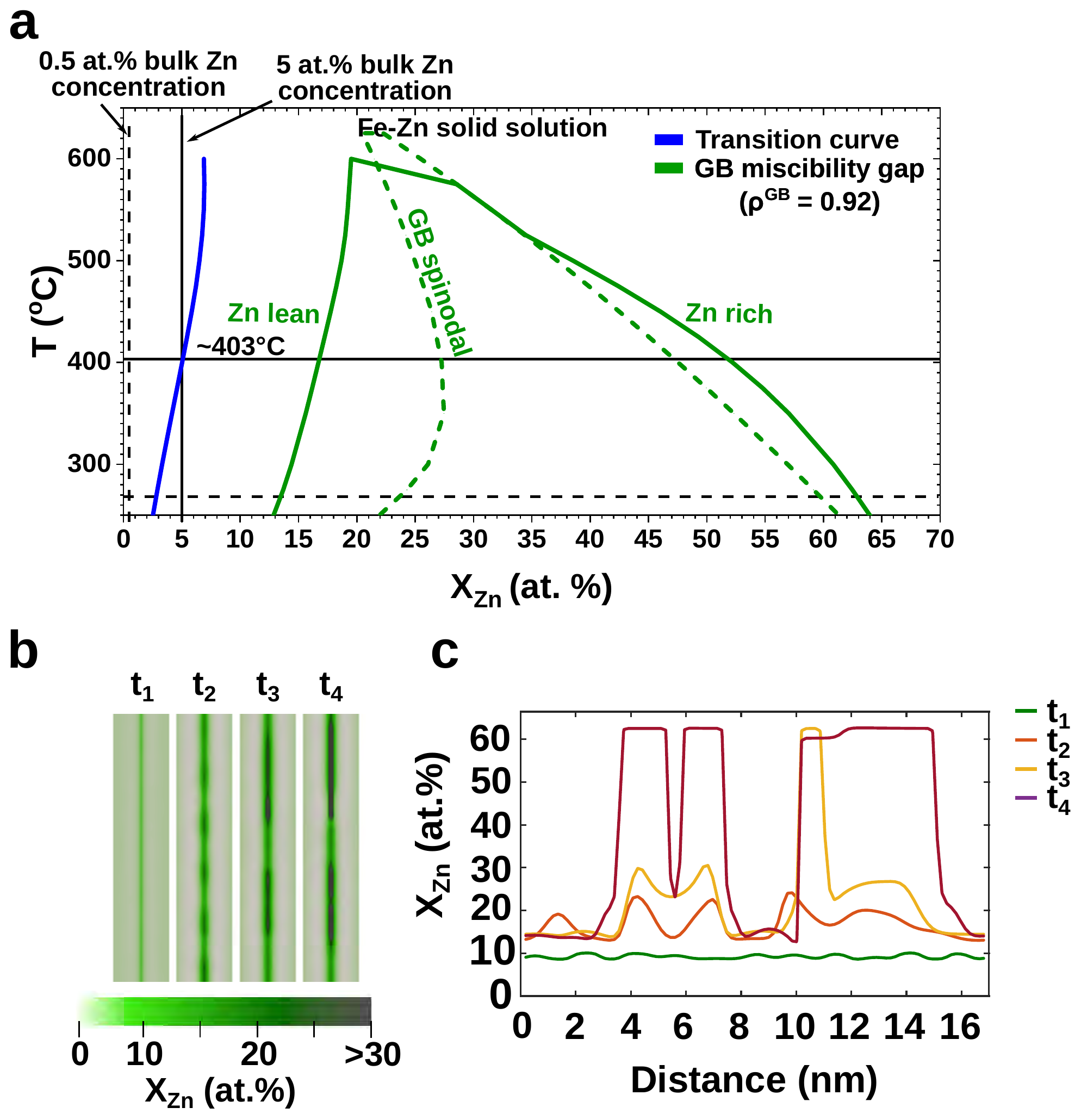}
\caption{\textbf{Density-based GB phase diagram and phase-field simulation.} \textbf{a} GB phase diagram for the Fe-Zn system showing a miscibility gap separating Zn-lean and Zn-rich GB phase regions. The solid curves represent the binodal, the dashed curves the spindal curves, respectively. The transition curve indicates the critical temperature and bulk Zn concentration above which the GB decomposes into Zn-rich and -lean domains. \textbf{b} Edge-on view of the time evolution of the GB phase decomposition for a bulk Zn concentration of 5 at.\% at 300$^\circ$C obtained by DPF simulations. The GB is oriented vertically and the scale bar is 5 nm. \textbf{b} Concentration profiles of Zn extracted from b) for each time step.}
\label{PF_SIM}
\end{figure*}

\vspace{1em}
\noindent{\textbf{Co-segregation behavior of Zn and B}}\\

\vspace{-1em}
In order to explore the interaction of B and Zn and its impact on the segregation behavior and GB cohesion, we performed first principles DFT calculations of the $\Sigma 5\,(3\,1\,0)$ as well as $\Sigma 5\,(2\,1\,0)$ GBs. Details on the computational approach can be found in the Methods section. The GB structures with all possible segregation sites considered for Zn (green) and B (blue) atoms are displayed in Fig.~\ref{pub2_DFT}~a and b. The same calculations performed for C (brown) are shown in the Extended Data Figure~\ref{DFT2_C}. The segregation energies of Zn $E_{seg}^{Zn}$ with and without interstitial B for the $\Sigma 5\,(3\,1\,0)$ and $(2\,1\,0)$ GBs are shown in Fig.~\ref{pub2_DFT}~c and d. Zinc alone has a similarly high segregation energy for its lowest energy position at the GB center ($z=0~\AA$) in a substitutional site within the GB plane for both GBs of $-0.57\,eV$ and $-0.58\,eV$, respectively. However, if B is present, the overall segregation energies of Zn are strongly reduced (see the cyan triangles). Note that multiple possible configurations were investigated. For example, the presence of B lowers the segregation energy of Zn at the GB center position by a factor of $\sim$6 to $-0.09\,eV$ for the $\Sigma 5\,(3\,1\,0)$ GB. Only for Zn farthest away from the GB considered here at $z=\pm2~\AA$, a slightly higher $E_{seg}^{Zn}$ of $-0.30\,eV$ is obtained for a single configuration. A similar trend is observed for the $\Sigma 5\,(2\,1\,0)$ GB, where the segregation energies are reduced by a factor of $\sim$2-3 in the presence of B. A possible configuration with Zn at the GB center with nearly the same segregation energy as without B is found. This suggests that the $\Sigma 5\,(2\,1\,0)$ GB shows a slightly higher probability for Zn to segregate with B being present than the $\Sigma 5\,(3\,1\,0)$ GB. 

Based on the calculated segregation energies, we estimate the enrichment of Zn at the GB at $800^\circ$C using the White-Coghlan isotherm \cite{whiteSpectrumBindingEnergies1977} considering segregation to all sites at the GB. This allows us to explore the influence of the bulk Zn concentration and the impact of B on the expected amount of Zn at the GB in the theoretical high temperature limit, which corresponds to the annealing temperature of the bicrystal diffusion couple. Since both the $\Sigma 5\,(3\,1\,0)$ and $\Sigma 5\,(2\,1\,0)$ GBs show a similar minimum segregation energy of Zn alone, the Zn concentration is predicted to be nearly the same at both GBs. It increases monotonically with increasing Zn bulk content as shown in Fig.~\ref{pub2_DFT}e (green dots). When B is also present at the GBs, a strong reduction of the maximum Zn content can be observed for both interfaces. For the $\Sigma 5\,(3\,1\,0)$ GB (solid lines) a reduction in the Zn concentration from $85\,at.\%$ to $30\,at.\%$ is observed for a bulk Zn concentration of $5.5\,at.\%$, which corresponds to the Zn-rich Region $1$ in the experiments. A similar trend is observed for lower Zn bulk contents. The reduction in GB Zn concentration is less pronounced for the $\Sigma 5\,(2\,1\,0)$ boundary as illustrated in Fig.~\ref{pub2_DFT}~e (dashed lines). These predictions indicate that B does not only strongly reduce the segregation energies, but with this also significantly lowers the maximum attainable Zn content at the GBs within the thermodynamic limit at the annealing temperature.

\vspace{1em}
\noindent{\textbf{Grain boundary cohesion}}\\

\vspace{-1em}
To assess the effects of co-segregation of Zn and B on GB cohesion, we calculated the theoretical work of separation $W_{sep}$ for the $\Sigma 5\,(3\,1\,0)$ and $\Sigma 5\,(2\,1\,0)$ GBs as shown in Fig.~\ref{pub2_DFT}~f. GB cohesive properties were computed for each Zn concentration in the most preferential segregation state. Both GBs show very similar cohesive properties in the pure state without any solutes present resulting in a work of separation of $3.5\,J/m^2$. With the addition of a single B atom, corresponding to half a monolayer (ML) or $\sim 9$~at.\% and $\sim 6$~at.\% for the $\Sigma 5\,(3\,1\,0)$ and $\Sigma 5\,(2\,1\,0)$ GB, respectively, the cohesive energy increases to $4.3$ and $4\,J/m^2$. We then investigated the effects of Zn additions to the GBs and found that the $W_{sep}$ decreases almost linearly for both GBs, albeit with a steeper slope for the $\Sigma 5\,(3\,1\,0)$ GB. For the case when no B is present at the GBs, four Zn atoms ($=2$~MLs) lower the work of separation to $2.3\,J/m^2$ for the $\Sigma 5\,(3\,1\,0)$ GB. The reduction of $W_{sep}$ is less pronounced for the $\Sigma 5\,(2\,1\,0)$ GB with $2$~ML Zn to $3\,J/m^2$. If both B and Zn are considered, it is found that the cohesion reducing effects of Zn are compensated by the presence of B. Even when 3 Zn atoms ($=1.5$~ML) are added to both GBs, the work of separation is still higher than that of the pure interfaces. This shows that the effect of impurity atoms on GB cohesion is to a large extent purely additive. 

From these observations clear implications can be established for GB-based alloy design to prevent LMIE. The addition of B and other impurity elements can reduce the tendency for Zn segregation, while the presence of cohesion enhancing elements such as B counteracts the embrittling tendency of Zn.

\begin{figure*}[htbp]
\includegraphics[width=0.98\linewidth]{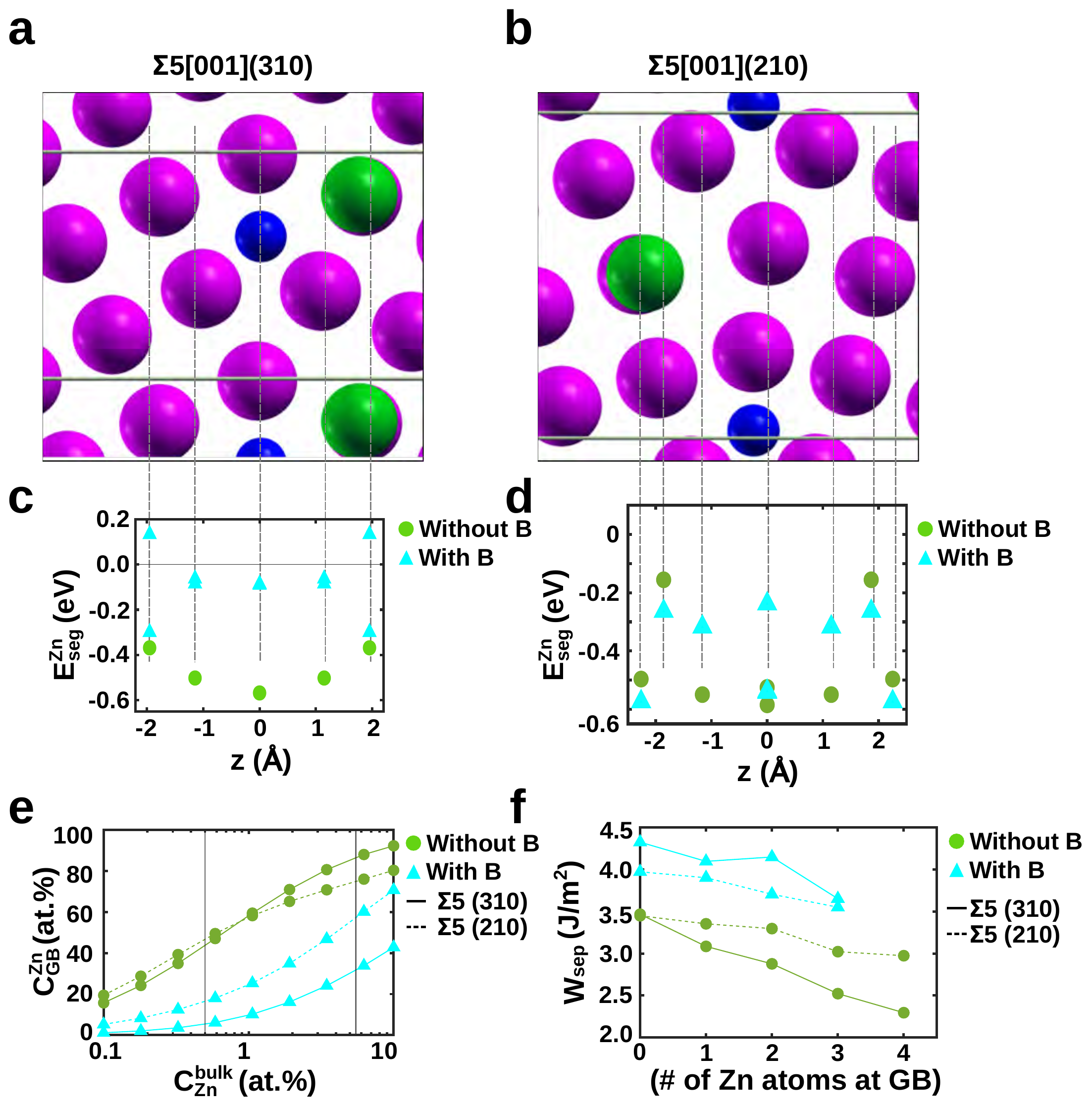}
\caption{\textbf{Modelling of Zn segregation to $\Sigma 5\,[0\,0\,1]$ GB with and without B.} Atomic structure of \textbf{a} $\Sigma 5\,[0\,0\,1](3\,1\,0)$ and \textbf{b} $\Sigma 5\,[0\,0\,1](2\,1\,0)$ GB with interstitial B (blue) and substitutional positions (purple = Fe, green = Zn). \textbf{c, d} Zn segregation energies from DFT for Zn to pure GB (green circles) and to GB when B is present (cyan triangles). \textbf{e} Zn concentration at the GB as a function of the Zn bulk concentration modelled for $\Sigma 5\,[0\,0\,1](3\,1\,0)$ (solid line) as well as $\Sigma 5\,[0\,0\,1](2\,1\,0)$ (dashed line) GB. For both GBs the enrichment level was calculated with (cyan) and without (green) B. The vertical lines highlight the observed Zn concentrations in bulk - namely $0.5\,at.\%$ and $5.5\,at.\%$. \textbf{f} Modelled GB cohesion for different number of Zn atoms at $\Sigma 5\,[0\,0\,1](3\,1\,0)$ (solid line) and $\Sigma 5\,[0\,0\,1](2\,1\,0)$ (dashed line) GB with (cyan) B and for GBs where B is not present (green).}
\label{pub2_DFT}
\end{figure*}

\section*{Discussion}
In this study, the nanoscale segregation behaviour of Zn in the presence of B and C at a $\Sigma 5\,(3\,1\,0) [0\,0\,1]$ bcc-Fe tilt GB was investigated. It is found that Zn is forming a periodic, line-type segregation pattern with nanometer periodicity in regions with $\sim$5~$at.\%$ Zn bulk concentration. Despite the low Zn bulk concentration, strong modulation of Zn-rich ($\sim$35~$at.\%$) and Zn-lean ($\sim$15~$at.\%$) regions along the GB were observed which originate from a phase separation of the GB during cooling from the segregation transition at $\sim$403$^{\circ}$C according to our thermodynamically informed phase-field simulations. However, we can also not exclude the role of GB defects in the formation of this Zn segregation pattern. These Zn modulations alone would lead to local fluctuations in the work of separation along the GB, given the results obtained by DFT. Such local compositional variations at GBs have not been taken into account to explain the origins of LMIE or GB fracture \cite{bealLiquidZincEmbrittlement2012, choMicrostructureLiquidMetal2014, ashiriLiquidMetalEmbrittlementfree2016, razmpooshLiquidMetalEmbrittlement2018, razmpooshRoleRandomCoincidence2020b, ikeda2022early, scheiberInfluenceAlloyingZn2020}.
From APT it is observed that the impurity elements B and C are homogeneously distributed within the GB plane in the same regions. However, their peak concentration is reduced compared to the as-grown bicrystal \cite{ahmadianAluminumDepletionInduced2021}. The asymmetric segregation profiles of B and C indicate that the GB has migrated during the annealing treatment either induced by Zn segregation \cite{hillertChemicallyInducedGrain1978, rabkinHighTemperatureDiffusionInducedGrain1993} or through the triple line, where the GB is connected to the liquid Zn reservoir \cite{dohie2007grain}.

Following our first-principles calculations, the strong repulsive interactions between B and Zn lead to a significant reduction in the segregation tendency of Zn in the presence of B. This ultimately leads to a reduction in the maximum Zn concentration at the GB, which by itself has a positive effect on GB cohesion \cite{tahir2014hydrogen, scheiberImpactSoluteSoluteInteractions2018}. Furthermore, the trends in calculated cohesive properties of the GB clearly indicate that even lower levels of B comparable to those observed experimentally are capable to compensate for the embrittling effects of Zn. Since both B and C are known to enhance GB cohesion \cite{ahmadianAluminumDepletionInduced2021, wuEffectsCarbonFeGrainBoundary1996, fraczkiewiczInfluenceBoronMechanical2000, wangFirstPrinciplesStudyCarbon2016, miyazawaAtomicBondbreakingBehaviour2017}, our calculated values for the work of separation only considering B are a lower bound and it is shown in the Extended Data that C has a similarly positive effect on GB cohesion.

While LMIE is a complex, hierarchical process leading to material failure by the ingress of liquid metal along GBs, it has been indicated that GB composition and local co-segregation effects ahead of the crack tip are decisive in the underlying mechanisms \cite{razmpooshRoleRandomCoincidence2020b, digiovanniLiquidMetalEmbrittlement2021}. Our investigations show that the evolving GB composition and intrinsic elemental interactions are crucial in understanding their effects on GB cohesion. 

Our work emphasizes the role of cohesion enhancing impurity elements and how they can be used to prevent GB failure induced during liquid metal embrittlement. We demonstrate that impurity segregation can mitigate the embrittling effects of Zn by hindering its segregation to GBs and leveling its reduction in GB cohesion. This points toward a promising route for materials design in which the controlled addition of B and C can act as efficient agent for reducing Zn mediated LMIE in steels. 

\section*{Methods}
\subsection*{Experimental}
For the experiments, a Fe-$4\,at.\%$Al bicrystal with a global $\Sigma 5\,[0\,0\,1]\,(3\,1\,0)$ was grown with the in-house Bridgman technique. The bicrystal was cut into a $\sim 15 \times 11 \times 1\,mm$ rectangle with the GB in the center. The steps how the bicrystal was coated with Zn are shown the Supplementary Figure 1. First, the sample was mechanically grinded and then chemically polished using a solution of $6\%$ HF, $14\%$H$_2$O and $80\%$H$_2$O$_2$. In order to prevent the contamination of the sample after polishing, the sample was held with steel rods. Afterwards, the sample was immersed into a $99.999\%$ Zn bath for $300\,s$. The temperature of the Zn bath was kept at $\approx 467^\circ C$. The Zn was then grinded off the large $(0\,0\,1)$ surfaces as well as the large sides of the sample. Finally the sample was encapsulated in a quartz tubes under vacuum ($\sim 10^{-3}-10^{-4}\,mbar$) and annealed at $800^\circ C$ for $80$ hours. The Zn diffused bicrystal was mechanically polished until reaching a mirror-like surface. Microstructural characterization of the GB was done by EBSD (see Supplementary Figure 2a) as well as EDS, which were simultaneously carried out by means of a ThermoFisher Scios 2 dual beam FIB/SEM equipped with an EDAX Velocity EBSD camera and an EDAX Octane Elite Super EDS detector. From the characterized area, TEM plane-view specimen were prepared from two different regions using the Scios 2 FIB. STEM imaging as well as STEM-EDS were performed on a C$_s$ probe-corrected FEI Titan Themis $60-300\,kV$ operated at $300\,kV$ as well as $120\,kV$ equipped with the SuperX EDS detector system. The investigations were done with a semiconvergence angle of $17\,mrad$, a camera length of $100\,mm$ and a probe current of $80\,pA$. The high resolution images were processed with a Butterworth filter and a Gaussian filter. The APT experiments were done in a CAMECA LEAP 5000 XS operated in laser-mode.

\subsection*{Computational Details}

\subsubsection*{Atomistic Simulations}
The computational details are described in more detail in Ref.~\cite{ahmadianAluminumDepletionInduced2021,scheiberImpactSoluteSoluteInteractions2018} but encompass first principles density functional theory using the Vienna Ab-initio Simulation Package (VASP) with projector augmented wave-functions~\cite{kresseInitioMolecularDynamics1993,kresseEfficiencyAbInitioTotal1996,kresseEfficientIterativeSchemes1996,kresseUltrasoftPseudopotentialsProjector1999}. For the exchange-correlation part, the PBE functional~\cite{perdewGeneralizedGradientApproximation1996,perdewPerdewBurkeErnzerhof1998} was employed. Other parameters include the k-point mesh set as close as possible to 40~k-points/\AA{} for all involved simulation cells, and the energy cutoff with 400~eV. The GBs for simulations are constructed from two symmetrically misoriented slabs that meet at the GB on one side and on the other side are separated by a vacuum layer of at least 8~\AA{}. Segregation energies are computed using \cite{ebner2021grain}
\begin{equation}
E_{seg,i}^x = \left(E_{gb,i}^x - E_{gb}\right) - \left(E_{bulk}^x - E_{bulk}\right) + \delta_{i-s} E_{Fe},
    \label{eq:eseg}
\end{equation}
with $E_{gb}$ being the total energy of a GB  without solute $x$ while $E_{gb,i}^x$ denotes the total energy of the same cell but with the solute $x$ at GB site $i$. For the reference of the solute in the bulk, the difference is computed between a bulk cell of 128 atoms without solute $x$ and with solute $x$, $E_{bulk}$ and $E_{bulk}^x$, respectively. For considering the change of interstitial B in the bulk to a substitutional site at the GB, $\delta_{i-s}=1$ and $E_{Fe}$ corresponds to the total energy of a single Fe atom in its bulk structure . Segregation energies are connected to the solute enrichment at GBs via the White-Coghlan segregation isotherm~\cite{whiteSpectrumBindingEnergies1977} extended to multiple solutes~\cite{scheiberSoluteDepletionZones2020}. 

The work of separation was computed by separating the GB at all possible GB planes by at least 8~\AA{} and computing the difference of the total energies for separated and joined GB slabs in the simulation cell \cite{scheiberImpactSoluteSoluteInteractions2018,scheiberInfluenceAlloyingZn2020a}. The lowest obtained work of separation for each segregated state is considered as the most probable fracture plane.

\subsubsection*{Phase-field Simulations}
The composition- and temperature-dependent GB phase behavior as well as GB segregation evolution are explored using the thermodynamically-informed phase-field modelling. The derivation and computational details of the model can be found in \cite{kamachaliModelGrainBoundary2020a}. The CALPHAD integrated density-based free energy functional of the Fe-Zn system reads
\begin{align}
\ & G(\rho, X_{Zn}, T)  = X_{Fe} G_{Fe} (\rho, T) + X_{Zn} G_{Zn} (\rho, T) \nonumber \\
& + \rho^2 \Delta H_{mix}^B (X_{Zn}, T) - T \Delta S_{mix}^B (G_{Zn}, T) \nonumber \\
& + \frac{\kappa_{X_{Zn}}}{2} (\nabla X_{Zn})^2
\label{G_alloy}
\end{align}
with $G_i(\rho, T) = {\rho^2} E_i^B+{{\rho}} (G_i^B (T) - E_i^B) + {\frac{\kappa_i^\rho}{2} \left( \nabla \rho \right)^2}$ and the atomic density $\rho \in [\rho^{GB}, 1]$ ($\rho = 1$ for the bulk and $\rho = \rho^{GB}$ at the GB plane). Here the superscript $B$ indicates bulk properties extracted from TCFE11 Thermo-Calc database: $G_i^B$ is the free energy, $E_i^B$ the potential energy and $\Delta H_{mix}^B$ and $\Delta S_{mix}^B$ are the mixing enthalpy and entropy, respectively. The gradient energy coefficients $\kappa_i^\rho$ and $\kappa_{X_{Zn}}$ accounts for the structural and chemical heterogeneity with the GB region. 

The GB phase diagram is calculated for $\rho^{GB} = 0.92$ which corresponds to the current $\Sigma 5\ [0\,0\,1]$ GBs. Figure \ref{suppFig_DPF1} shows the density profiles across the GBs computed from previous atomistic simulations \cite{ratanaphan2015grain}. The chemical spinodal and miscibility gaps are computed using a robust implementation of the Maxwell construction and verified versus Thermo-Calc calculations. The density-based phase-field simulations were performed using an Openmp parallel C++ code with adaptive time-stepping. The simulation domain has been 17x17x3.4 nm$^3$ with dx = 0.17 nm. Further details of simulation methods can be found in \cite{darvishikamachali2020segregation}. 

\section*{Data Availability}
The data that support the findings of this study are available from the corresponding author upon reasonable request. 

\section*{Code Availability}
The code that support the findings of this study are available from the corresponding author upon reasonable request. 
\newpage
\beginsupplement
\section*{Extended Data}
\vspace{1em}
\noindent{\textbf{Diffusion couple}}\\

\vspace{-1em}
The steps to introduce Zn into the GB is shown in Fig.~\ref{suppFig1}. First, the bicrystal was mechanically grinded with SiC paper upon a grit size of $15~\mu m$. Next, the bicrystal was chemically polished with a acidic solution of $6\%$ HF, $14\%$ H$_2$O and $80\%$ H$_2$O$_2$. In order to avoid contamination, the sample was clamped by a steel rod. Immediately after the polishing step, the bicrystal was immersed into a Zn bath, whose temperature was measured to $467^\circ C$. The bicrystal was kept for $300\,s$ into the bath. Afterwards, the Zn layers onto the (001) surfaces and the large edges were mechanically removed, so that only Zn at the short edges remained. Under this condition, the Zn coated bicrystal was encapsulated in a quartz capsule under vacuum with $p \sim 10^{-3} \text{to} 10^{-4}\,mbar$ and annealed at $800^\circ C$ for $80$ hours. A backscattered electron (BSE) micrograph from the region below the Fe-Zn interface shown in Fig.~\ref{suppFig2}a reveals the formation of a two-phase layer containing $\alpha-$~Fe and the solid-solution phase of Zn. The orientation of the two-phase layer was determined by electron backscatter diffraction (EBSD) having an orientation close to $[0\,1\,2]$. The boundary type was identified as a symmetric $40^\circ \Sigma 5\,(3\,1\,0)[0\,0\,1]$ high angle GB. STEM investigations of the two phase region are shown in Fig.~\ref{suppFig2}b and c. Figure~\ref{suppFig2}b shows the ABF-STEM image and the corresponding STEM-EDS elemental maps for Fe and Zn around the GB. As it can be seen, the GB is dewetted with Zn. A concentration profile across the dewetted boundary results in a Zn concentration of $\geq 60\,at.\%$ (see Figure~\ref{suppFig2}c).
\begin{figure*}[htbp]
\centering
\includegraphics[width=0.9\linewidth]{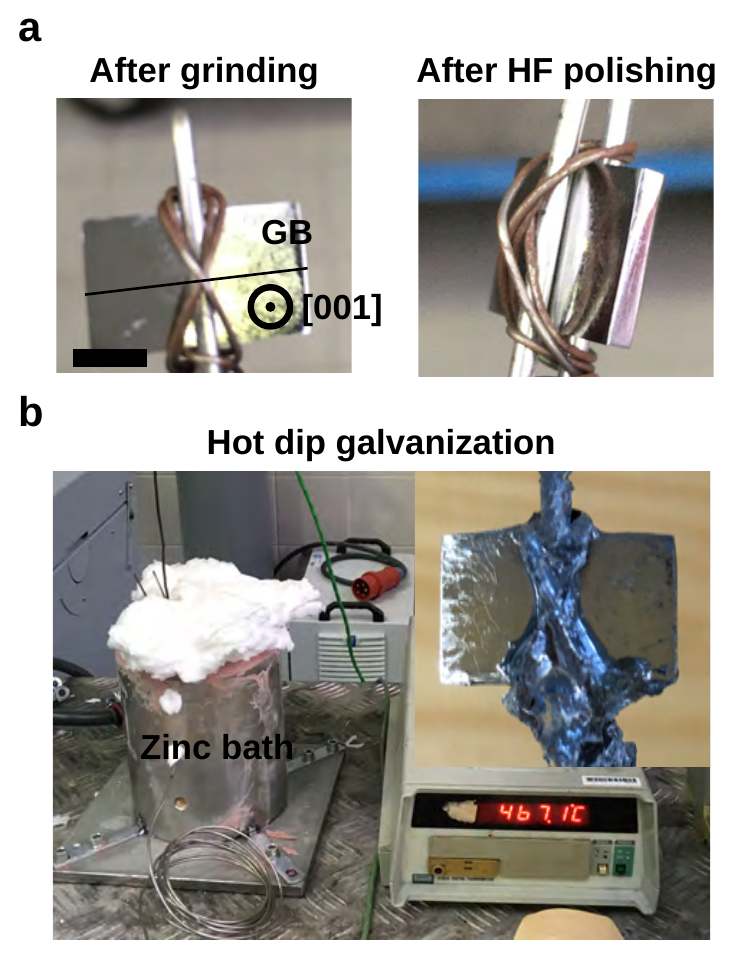}
\caption{\textbf{Zn diffusion into the Fe bicrystal.} \textbf{a} Mechanically grinding the bicrystal up to $1200$ SiC grit paper and subsequent chemical HF polishing. The GB is indicated by a black line. \textbf{b} The polished sample was immersed into a liquid Zn bath. The temperature of the bath was measured with a thermocouple to be $467^\circ C$. The inset shows the bicrystal completely coated with Zn.}
\label{suppFig1}
\end{figure*}

\begin{figure*}[htbp]
\centering
\includegraphics[width=.7\textwidth]{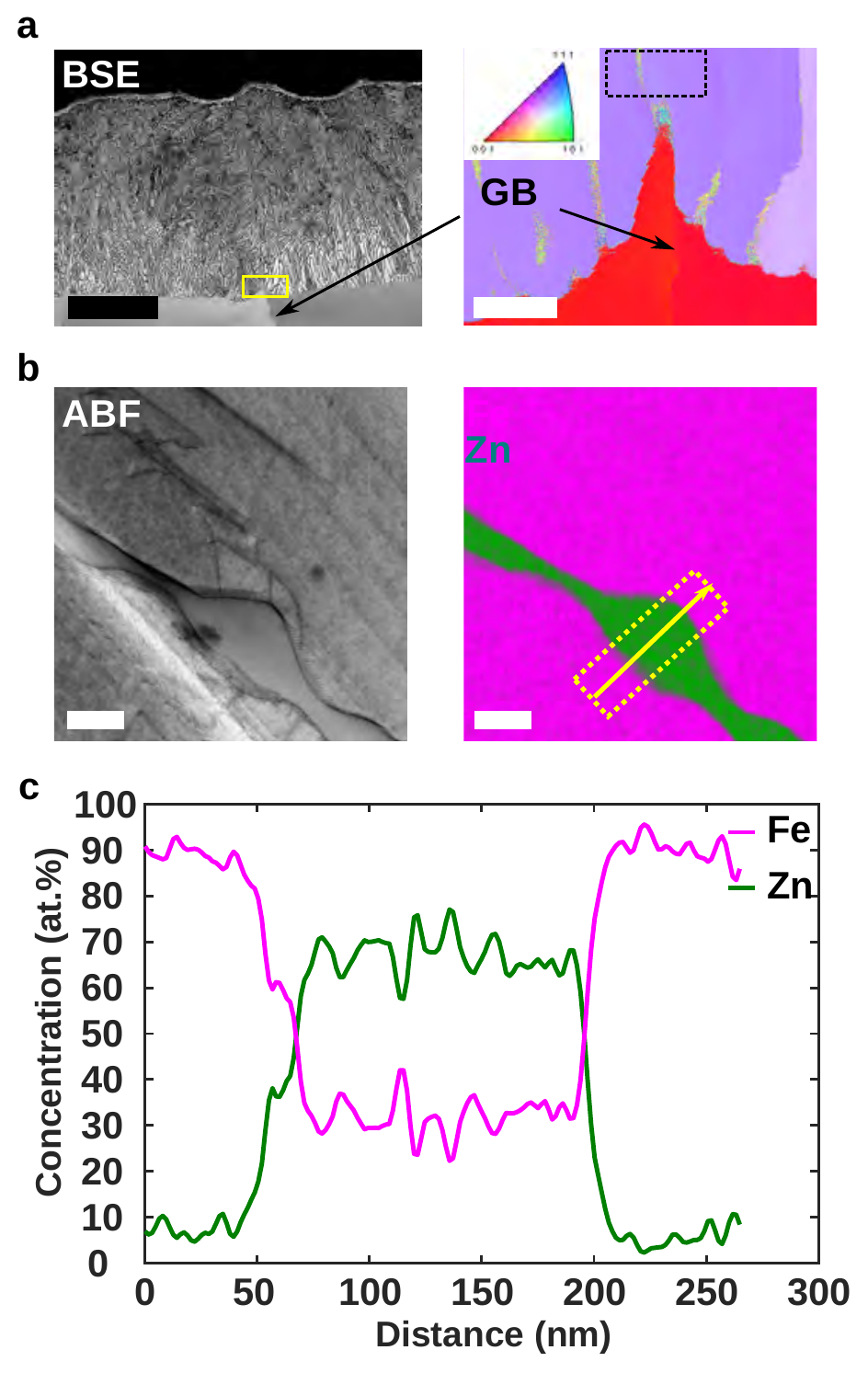}
\caption{\textbf{Microstructural characterisation} \textbf{a} A BSE image shows the formation of a two-phase region, which extended up to $\sim 150\,\mu m$. The out-of plane inverse pole figure shows orientation of the two-phase region to be close to $[0\,1\,2]$, while the $\alpha -$Fe region maintained the $[0\,0\,1]$ direction. TEM specimen were extracted from the black dashed box. \textbf{b} ABF-STEM image and corresponding STEM-EDS mapping of the GB show the GB completely dewetted and transformed into a Zn-rich phase. The line profile in \textbf{c} shows a concentration of $\sim 70\,at.\%$ Zn. The scale bars in a) are $50\,\mu m$ (left) and $2\,\mu m$ (right) and in \textbf{b} it is $100\,nm$}
\label{suppFig2}
\end{figure*}

\vspace{1em}
\noindent{\textbf{Influence of temperature on GB structure}}\\
STEM investigations of the GB showed mesoscale inclinations with symmetric $(2\,1\,0)$ and asymmetric segments. In order to understand whether the high temperature annealing was causing it, a Zn-free bicrystal was annealed at $800^\circ C$ for $80$ hours under similar high vacuum conditions ($p \sim 10^{-5}\,mbar$). As shown in Fig.~\ref{suppFig5}a and b, the GB is flat and straight with no inclinations or kinks. The ABF-STEM shows dislocations nucleated within the grains, which can be caused during FIB milling. High-resolution HAADF-STEM images at different regions of the GB are shown in Fig.~\ref{suppFig5}c. The GB is a symmetric $\Sigma 5\,[0\,0\,1]\,(3\,1\,0)$ and its structure shows kite-type structural units similar as investigated in previous studies \cite{ahmadianAluminumDepletionInduced2021}. This confirms that the annealing does not affect the orientation of the boundary planes. Therefore, the observed inclinations needs to be related to the diffusion of Zn and the formation of the two-phase layer.
\begin{figure*}[htbp]
\centering
  \includegraphics[width=0.6\linewidth]{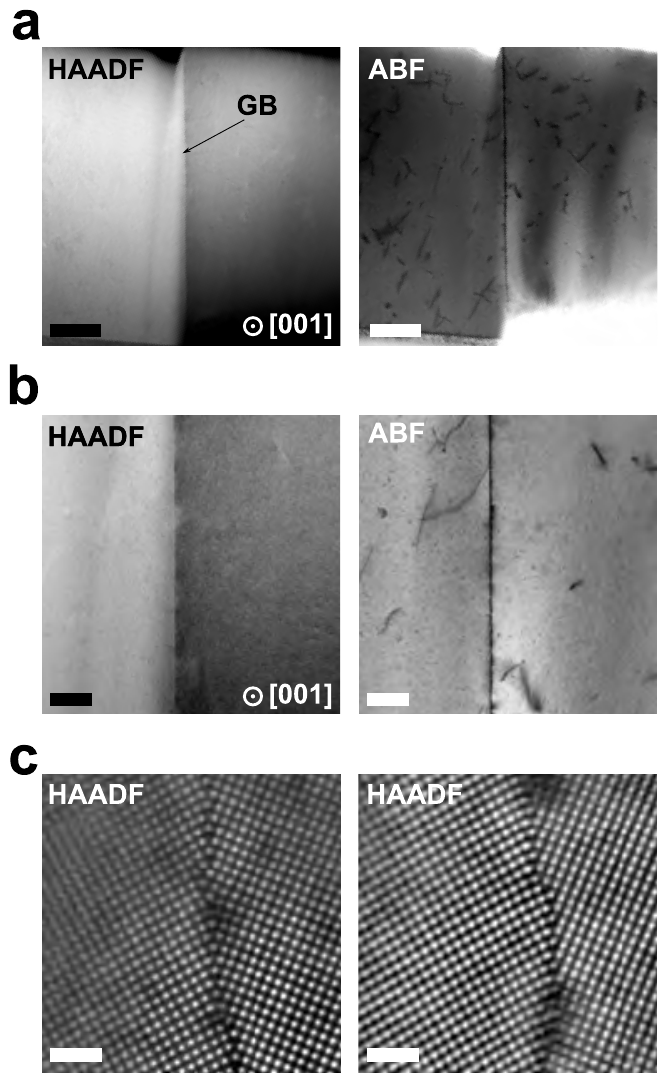}
\caption{\textbf{Influence of annealing on GB structure} \textbf{a} HAADF as well as ABF-STEM overview images of the GB show no inclinations of the GB, but the nucleation of many dislocations. A magnified image confirms that the GB did incline at the mesoscale (\textbf{b}). \textbf{c} Atomic-resolved HAADF-STEM images show the kite-type SU of the GB with a symmetric $(3\,1\,0)$ boundary plane. The GB contains few defects. The Scale bar in \textbf{a} is $500\,nm$, \textbf{b} it is $100\,nm$ and in \textbf{c} it is $1\,nm$.}
\label{suppFig5}
\end{figure*}
\vspace{-1em}

\begin{figure*}[htbp]
\centering
  \includegraphics[width=\linewidth]{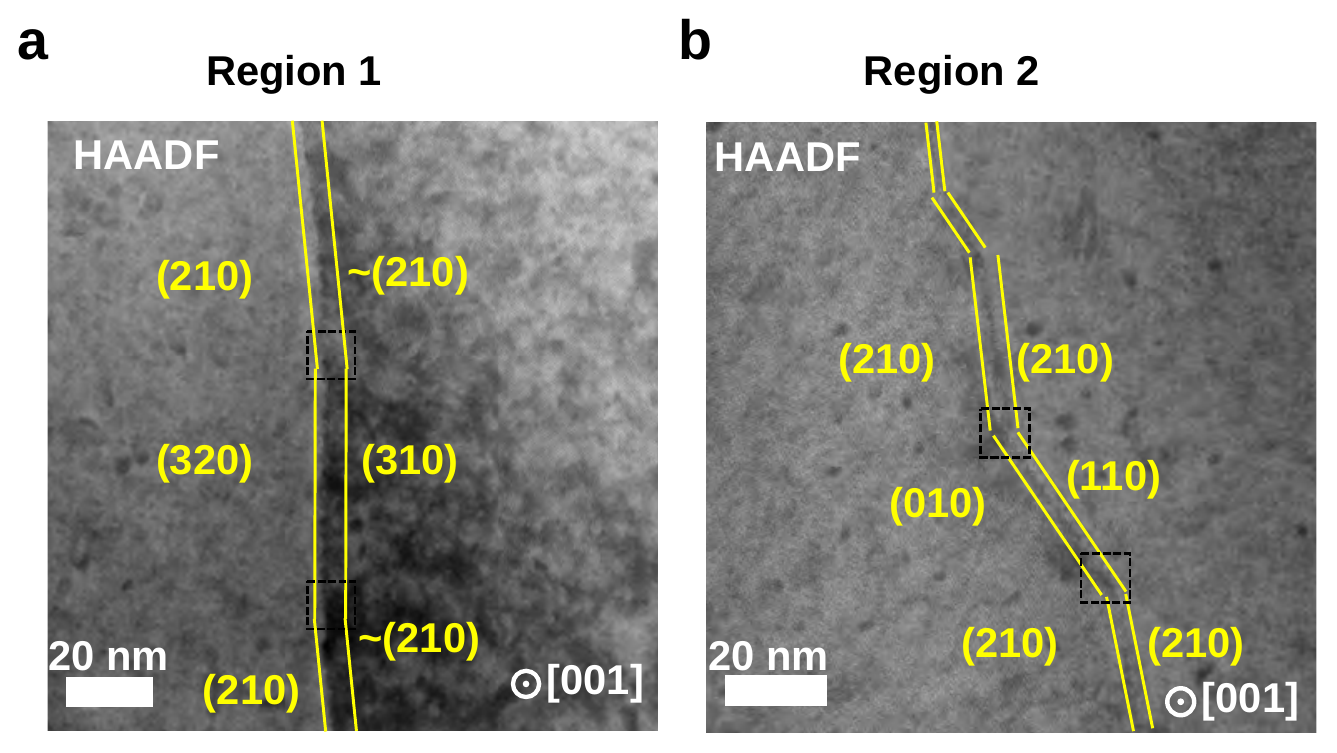}
\caption{\textbf{HAADF-STEM of the $\Sigma 5\, [0\,0\,1]$ GB from Region 1 and 2.} \textbf{a} Overview HAADF-STEM image showing the inclinations of the GB from nearly symmetric $(2\,1\,0)$ into an asymmetric configuration of $(3\,2\,0)/(3\,1\,0)$ boundary planes. The dashed square indicates the region analyzed at atomic level shown in Fig.~\ref{pub2_HRSTEM}a and b. \textbf{b} Overview HAADF-STEM image of the GB from the Region 2. The GB inclinates from symmetric $(2\,1\,0)$ into an asymmetric configuration of $(0\,1\,0)/(1\,1\,0)$ boundary planes. The Scale bar in \textbf{a} and \textbf{b} is $20\,nm$.}
\label{suppFig3}
\end{figure*}

\begin{figure*}[htbp]
\includegraphics[width=0.98\linewidth]{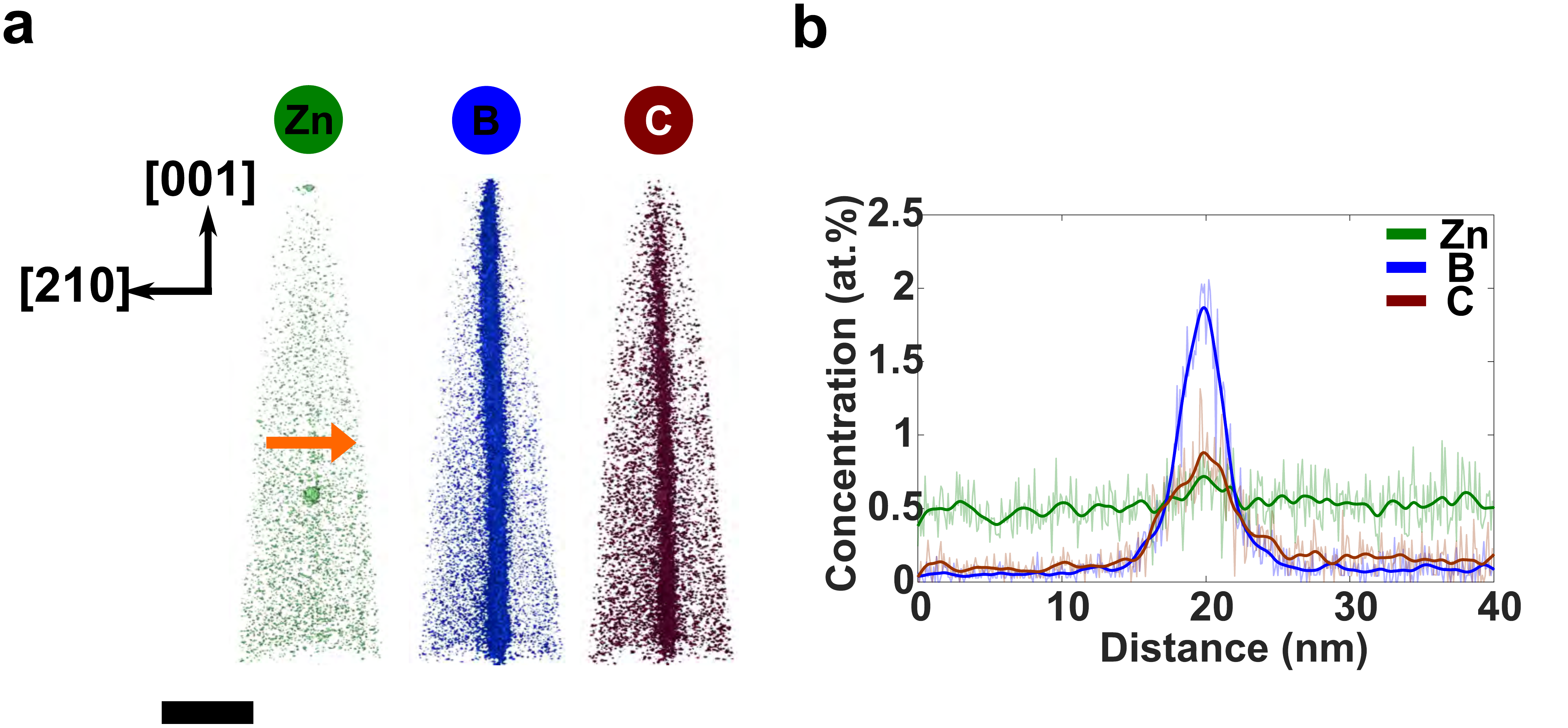}
\caption{\textbf{Segregation of Zn, B, and C at the $\Sigma 5\,(2\,1\,0) [0\,0\,1]$ GB of Region $2$.} \textbf{a} $3$D APT reconstruction showing the distribution of Zn, B and C. The $[0\,0\,1]$ tilt axis shows upwards. Using an isoconcentration value of $0.5\,at.\%$ for Zn and B and $0.3\,at.\%$ for C highlights the segregation of these elements to the GB. \textbf{b} The composition profile extracted from a cylindrical region with diameter of $30\,nm$ is extracted across the GB (shown as orange arrow in \textbf{a}). A clear increase of B and C is shown at the GB, while the concentration of Zn stays unchanged. The maximum concentration peak of B and C is at the same position of $\sim 18.5\,nm$. Both elements show a symmetric concentration profile. The scale bar in \textbf{a} is $40\,\text{nm}$.}
\label{pub2_APT2}
\end{figure*}

\begin{figure*}[htbp]
\centering
  \includegraphics[width=0.98\linewidth]{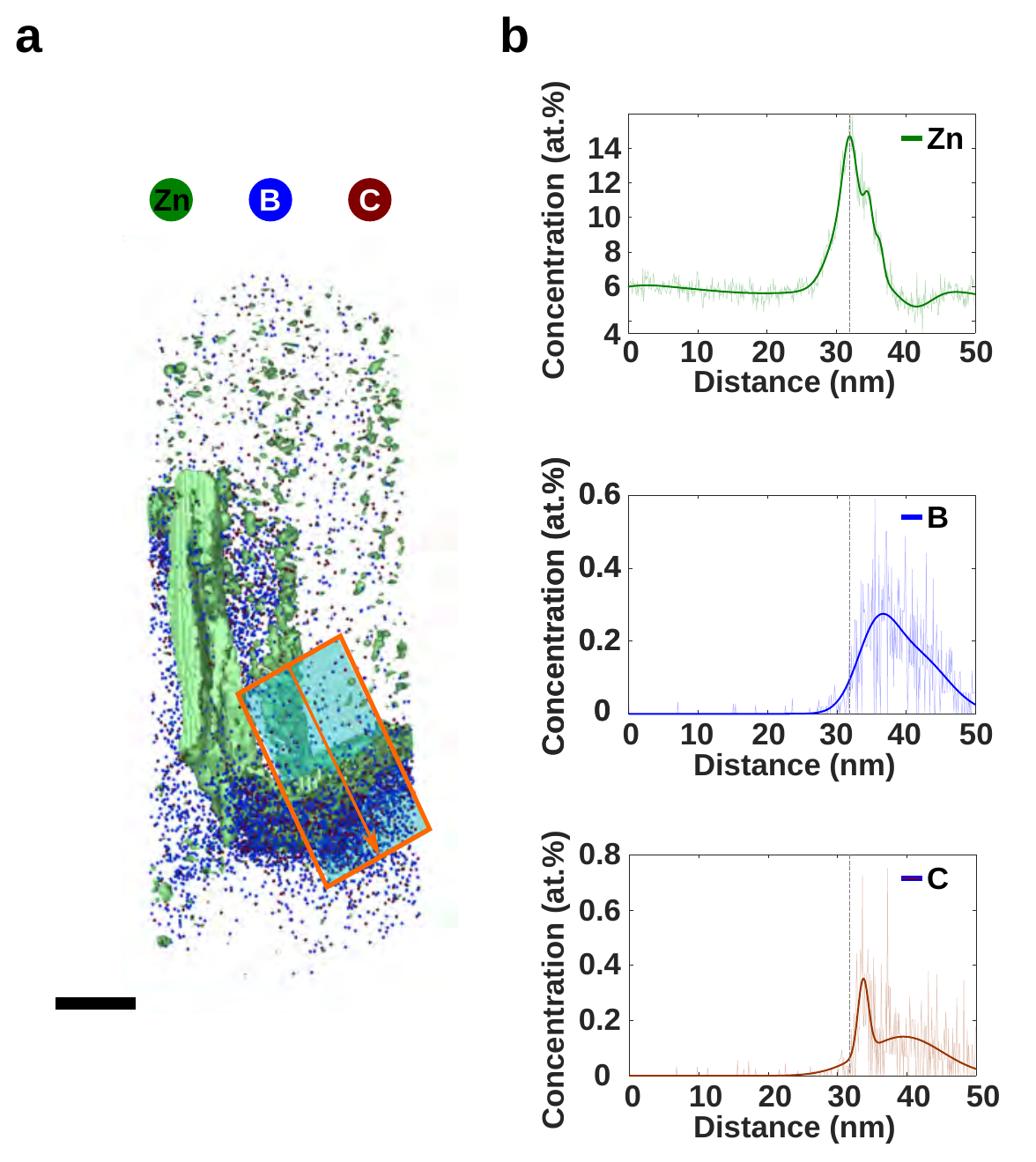}
\caption{\textbf{Segregation of Zn, B, and C at the $\Sigma 5\, [0\,0\,1]$ GB of Region 1}. \textbf{a} 3D APT reconstruction of showing the segregation of Zn, B and C at the GB. The GB is split into two parts connected through a large step, which is also enriched with Zn. \textbf{b} A concentration profile across the bottom GB is shown. Strong segregation of Zn and a moderate enrichment level of B and C of $\leq 0.5\,at.\%$. The Zn distribution shows a stronger depletion into the right grain, where the concentration peak of B and C are located. The latter elements show a tail into the right grain. The scale bar in \textbf{a} is $20\,nm$.}
\label{suppFig4}
\end{figure*}

\begin{figure*}[htbp]
\centering
  \includegraphics[width=\textwidth]{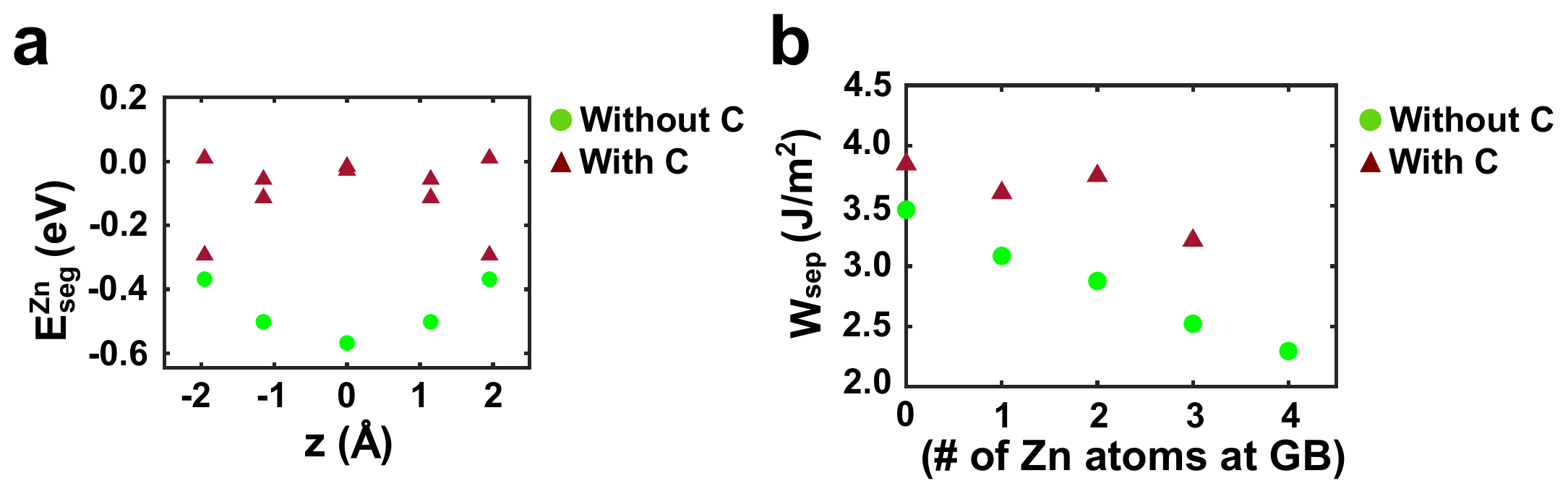}
\caption{\textbf{Modelling of Zn segregation to $\Sigma 5\, [0\,0\,1]\,(3\,1\,0)$ GB with and without C} \textbf{a} Zn segregation energy from DFT for Zn to pure (green circles) and to GB when C is present (brown triangles). The segregation energy of Zn and therefore its tendency decreases with the presence of C. \textbf{b} Calculated work of separation of the boundary for different number of Zn atoms with (brown) and without C (cyan). The presence of C shows an enhancement of the boundary strength.}
\label{DFT2_C}
\end{figure*}

\begin{figure*}[htbp]
    \centering
    \includegraphics[width=\textwidth]{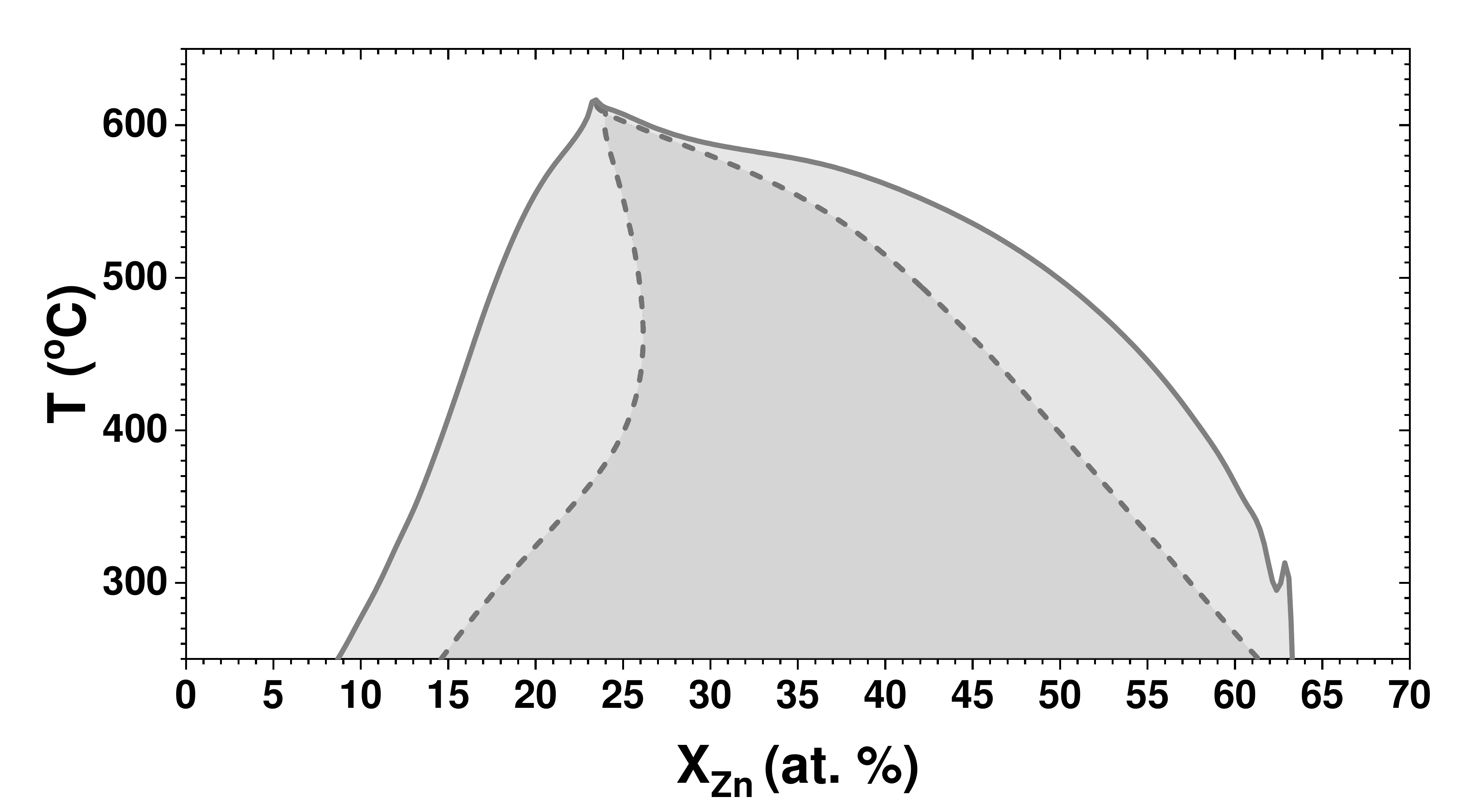}
    \caption{\textbf{Fe-Zn bulk phase diagram.} Bulk phase diagram of Fe and Zn determined by CALPHAD-based thermodynamic modelling. The shaded gray region represents the miscibility gap due to magnetic ordering.}
    \label{DPF_bulk}
\end{figure*}

\begin{figure*}[htbp]
\centering
  \includegraphics[width=\linewidth]{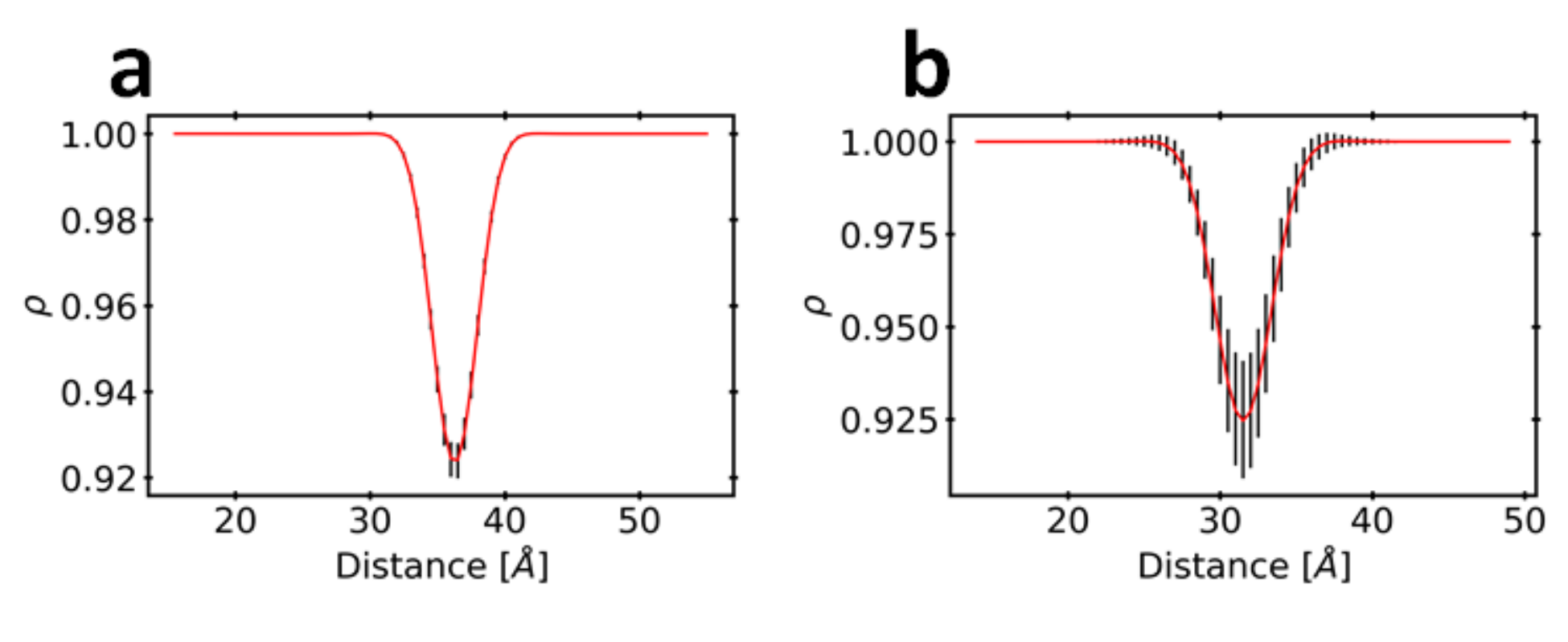}
\caption{\textbf{GB density profiles} atomic density and its fluctuations across the \textbf{a} $\Sigma 5\,[0\,0\,1] (3\,1\,0)$ and \textbf{b} $\Sigma 5\,[0\,0\,1] (2\,1\,0)$ GBs in $\alpha$-Fe. 3D atomic simulation results from \cite{ratanaphan2015grain} are used to compute the density profiles. All values are coarse-grained and averaged in planes parallel to the GB plane. The bars show the density fluctuation in the planes.}
\label{suppFig_DPF1}
\end{figure*}

\begin{figure*}[htbp]
\centering
  \includegraphics[width=\linewidth]{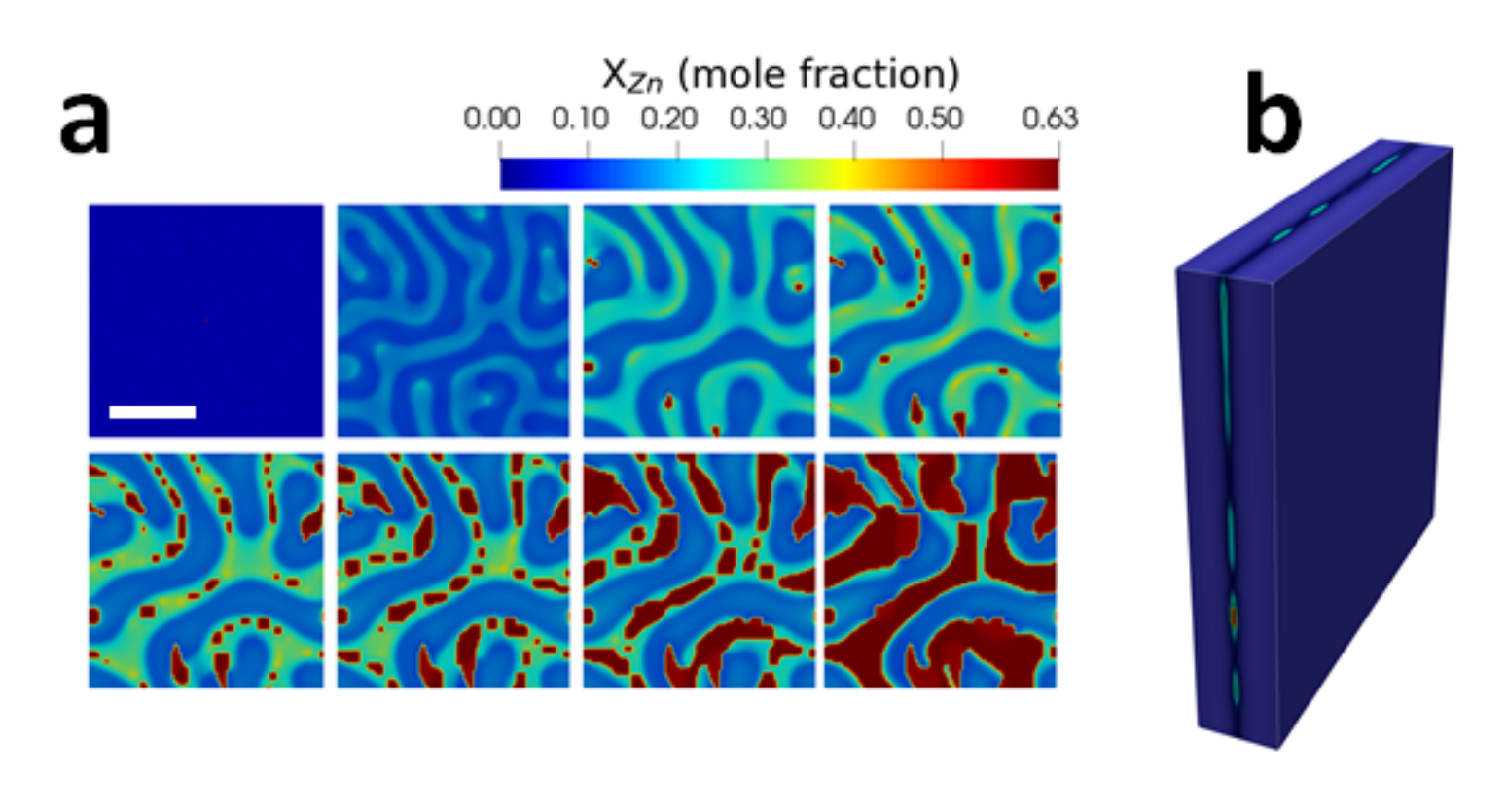}
\caption{\textbf{GB in-plane segregation and phase decomposition from DPF simulation} \textbf{a} the time evolution of Zn segregation and interfacial phase decomposition within the GB plane are shown. The scale bar is 4 nm. The results show the Zn segregation variation initiated by the GB density variations followed by a segregation transition and interfacial phase decomposition. \textbf{b} shows the 3D simulation domain.
}
\label{suppFig_DPF2}
\end{figure*}

\bibliographystyle{naturemag}
\bibliography{pub2}


\section*{Acknowledgments}
The authors gratefully acknowledge the financial support under the scope of the COMET program within the K2 Center “Integrated Computational Material, Process and Product Engineering (IC-MPPE)” (Project No 886385). This program is supported by the Austrian Federal Ministries for Climate Action, Environment, Energy, Mobility, Innovation and Technology (BMK) and for Digital and Economic Affairs (BMDW), represented by the Austrian Research Promotion Agency (FFG), and the federal states of Styria, Upper Austria and Tyrol. X. Z. is supported by the Alexander-Humboldt-Stiftung. R. D.K. gratefully acknowledges support from German Research Foundation (DFG) through projects DA 1655/2-1 and DA 1655/3-1. G. D. gratefully acknowledges support by the ERC Advanced Grant GB-Correlate (Grant Agreement 787446).

\section*{Author contribution}
G. D., W. E. and L. R. secured funding. C.H. L. and A. A. designed the experiments and C.H. L. supervised the project. A. A. conducted the electron microscopy experiments including STEM and FIB/SEM.\\ X. Z. conducted the APT experiments. D. S. performed the atomistic simulations. R. D. K. executed the thermodynamic and phase-field simulations. C.H. L., L. R. and B. G. contributed to the data analysis. A. A., D. S. and C.H. L. wrote the paper. X. Z., B. G., R. K. D., W. E., L. R. and G. D. revised the paper.

\section*{Competing interests}
The authors declare no competing interests.
\end{otherlanguage}
\end{document}